\documentclass[a4paper,11pt]{article}
\usepackage{jcappub}
\usepackage{aas_macros}

\newcommand{\nv}{\hat{\bf n}}
\newcommand{\bU}[1]{\langle bU_{\rm #1}\rangle}
\newcommand{\wtj}[6]{\left(\begin{array}{ccc} #1 & #2 & #3\\#4 & #5 & #6\end{array} \right)}

\title{The moving lens effect: analytical modelling and foreground suppression}

\author{Amy Wayland$^1$,}
\author{David Alonso$^1$,}
\author{and William Coulton$^1$}
\affiliation{$^1$Department of Physics, University of Oxford, Denys Wilkinson Building, Keble Road, Oxford OX1 3RH, United Kingdom}

\emailAdd{amy.wayland@physics.ox.ac.uk}

\abstract{The moving lens (ML) effect is a secondary anisotropy of the cosmic microwave background (CMB) generated by the transverse motion of gravitational potentials, providing a direct probe of the large-scale cosmic velocity field. Its detection relies on cross-correlating a CMB map with the transverse galaxy momentum field, constructed from the galaxy overdensity and a velocity field reconstructed from it. Restricting the reconstruction to large-scale modes strongly suppresses contamination from small-scale astrophysical foregrounds while preserving the ML signal. In this work, we develop a theoretical framework for the ML estimator and its foreground contamination. We show that the signal and foregrounds have a distinct dependence on the direction of the long-wavelength mode represented by the reconstructed galaxy velocity. In the squeezed limit enforced by the velocity reconstruction filter, the bispectrum sourcing the foreground correlation becomes independent of this direction, causing its leading contribution to vanish after angular averaging. In turn, the ML signal survives by matching this directional dependence, with its amplitude reduced only by the filtered velocity variance. Using the halo model, we derive the leading corrections beyond the squeezed limit and show that the residual contamination remains parametrically suppressed. Finally, we model the cross-correlation exactly in the curved sky, and show that it is sourced solely by the longitudinal component of the galaxy momentum field, in the form of a spin-1 $E$-mode, with all other contributions either strongly suppressed on small scales or exactly zero.}

\begin{document}
\maketitle
\flushbottom

\section{Introduction}
  The motion of large-scale structures relative to the cosmic microwave background (CMB) generates a secondary temperature anisotropy known as the moving lens (ML) effect \cite{1983Natur.302..315B}. Unlike gravitational lensing by static potentials, the ML effect is sourced by the transverse motion of gravitational potentials, which leads to an effective time variation in the gravitational well traversed by the CMB photons, akin to the integrated Sachs--Wolfe effect \cite{astro-ph/9501059,astro-ph/9803040}. The induced temperature fluctuation can be written as a line-of-sight integral that couples the velocity field $v_i$ to the gradient of the gravitational potential $\phi$,
  \begin{equation} \label{eq:lambda}
     \lambda(\nv) \equiv \left. \frac{\delta T}{T}\right|_{\rm ML} = -2 \int\mathrm{d}\chi\, v_i(\chi\nv)\, \partial_i\phi(\chi\nv).
  \end{equation}
  The ML effect therefore provides a direct probe of the large-scale cosmic velocity field. Because the effect is purely gravitational, it is not subject to the optical-depth degeneracy that limits the interpretation of kinetic Sunyaev--Zel'dovich (kSZ) measurements in terms of structure growth \cite{1812.03167,2108.02207}, and it is sensitive to the two transverse velocity components rather than the radial one, making the two probes naturally complementary. As upcoming CMB experiments are combined with increasingly deep and wide galaxy surveys, measurements of the ML effect are expected to become a powerful cosmological probe, providing new constraints on the growth of structure, tests of gravity on large scales, and information complementary to traditional peculiar velocity measurements \cite{1812.03167,2108.02207}.

  Traditional estimators of the ML effect are often based on directional stacking of CMB maps at the positions of sources oriented along the direction of their reconstructed transverse velocities and weighted by them \cite{2006.03060,2408.16055}. As shown in \cite{2605.15947}, this is equivalent to studying the harmonic-space cross-correlation of the CMB map with a map of the spin-1 reconstructed transverse galaxy momentum field (more concretely with its $E$-mode component). Other methods have also been proposed \cite{1812.03167,1812.04241}. These approaches remove any correlation with the statistically independent primary CMB fluctuations while retaining the ML signal. However, residual foreground emission, such as the thermal Sunyaev--Zel'dovich (tSZ) effect and the cosmic infrared background (CIB), traces the same underlying large-scale structure as the galaxy distribution and can therefore generate a correlated contribution to the estimator \cite{2305.15462,2401.12280,2408.16055}. Since the temperature fluctuations induced by the ML effect are extremely small, these foreground contaminants can easily dominate over the signal and have been identified as a key obstacle in detecting this effect. Interestingly, as shown empirically in \cite{Hotinli2026}, the application of a low-$k$ filter to the reconstructed velocity field of galaxies, isolating the large-scale velocity modes, drastically reduces the impact of foreground contamination on small angular scales, leaving the ML signal virtually untouched. This approach, combined with a harmonic-space analysis similar to that of \cite{2605.15947}, allowed \cite{Hotinli2026} to report the first detection of the moving lens effect.
  
  The intriguing behaviour of foregrounds in the presence of a Fourier-space low-pass filter was only demonstrated empirically in \cite{Hotinli2026}. While the effect may be understood intuitively in terms of separating the large-scale bulk motion from the velocity induced by nearby structures (which may drive the correlation with local foreground sources), to our knowledge no detailed theoretical study has been conducted to rigorously explain it. Likewise, while significant work has been devoted to the analytical modelling of the closely-related kSZ effect \cite{astro-ph/0106342,1109.0553,1301.3607,2208.07847,2306.03127,2509.18732}, the current treatment of the ML effect has so far been relatively simple, relying on the modelling of individual systems and the use of flat-sky and Limber approximations.
 
  In this paper, we address both of these shortcomings. We derive the moving-lens estimator in both the flat- and curved-sky limits, and develop a unified theoretical description of foreground contamination. The central observation is that the signal and the contamination enter the estimator through correlators of different order. The moving-lens signal is governed by a four-point function, whereas foreground contamination is sourced by a three-point function, leading to distinct angular behaviour. We demonstrate analytically that the leading foreground contribution vanishes in the squeezed limit, whereas the leading ML signal remains finite. The velocity filter therefore suppresses the ML signal only through an overall normalisation determined by the filtered velocity variance. We then derive the leading corrections beyond the squeezed limit using the halo model and show that the residual foreground contamination is parametrically suppressed.
 
  Finally, we generalise the formalism to the full sky using a spin-1 decomposition of the projected galaxy momentum field. We show that only the component of the momentum field longitudinal with respect to the Fourier wavevector contributes to correlations with scalar observables, so the transverse component decouples exactly on all scales. The full-sky formalism thus reproduces the flat-sky result in the large-$\ell$ limit while demonstrating that no significant large-scale leakage occurs.
 
  The structure of this paper is as follows. In Section~\ref{sec:ml}, we introduce the ML estimator and develop the flat-sky formalism, decomposing the projected galaxy momentum into components longitudinal and transverse to its Fourier wavevector, before performing the $E/B$ decomposition. In Section~\ref{sec:foregrounds}, we derive the foreground contribution to the estimator, show that its leading term vanishes in the squeezed limit while the ML signal remains finite, and derive the leading corrections beyond the squeezed limit within the halo model. In Section~\ref{sec:full_sky}, we generalise the formalism to the curved sky and establish its relationship to the flat-sky treatment. Section~\ref{sec:results} presents numerical evaluations of the halo model contributions to the foreground contamination, quantifies the suppression of foregrounds relative to the ML signal as a function of the velocity reconstruction scale, and compares the full calculations with the beyond-squeezed-limit predictions. Finally, we conclude in Section~\ref{sec:conclusions} with a summary of our main results and their implications for future CMB and galaxy surveys.
 
\section{The moving-lens estimator} \label{sec:ml}
 \subsection{The moving lens effect} \label{ssec:ml.definition}
  The peculiar velocity field, $v_i$, and the Newtonian potential, $\phi$, are related to the matter overdensity, $\delta_{\rm m}$, through the linearised continuity equation and the Poisson equation:
  \begin{equation}
      \mathbf{v}(\mathbf{k}) = \frac{i\mathbf{k}}{k^2} aHf \, \delta_v(\mathbf{k}), \quad\quad \phi(\mathbf{k}) = - \frac{3 H_0^2}{2ak^2} \, \Omega_{\rm m} \, \delta_{\rm m}(\mathbf{k}),
  \end{equation}
  where $f$ denotes the linear growth rate. We note that the linearised continuity equation is only accurate on large, linear scales. As we will see, only the large-scale velocity correlations enter the modelling of the ML effect considered here, and we therefore adopt the linear approximation for simplicity. We denote the matter overdensity sourcing the velocity field as $\delta_v$, in order to distinguish it from the full matter overdensity $\delta_{\rm m}$, and take it to be the \emph{linear} field $\delta_v\equiv\delta_{\rm L}$. This choice ensures that the velocity field contains no one-halo contribution in the correlators considered below. Importantly, this is a definition of the ML contribution adopted in our modelling, rather than an approximation to the full physical velocity field. Equation~\eqref{eq:lambda} describes the contribution to the time variation of the potential arising from its coherent translation with velocity $\mathbf{v}$. By contrast, internal, virialised motions within a halo induce an intrinsic time evolution of the potential rather than a rigid translation. We do not include these contributions in our definition of the ML effect; they may instead be viewed as part of the more general non-linear evolution of the potential associated with the Rees--Sciama effect. This distinction is partly conventional, and we return to the implications of neglecting these intra-halo velocity contributions in our conclusions. The impact of non-linearities in the velocity field has been studied in the context of the kSZ effect (see e.g. \citep{1506.05177}), and we leave a similar study regarding the ML effect for future work.
  
  Substituting these relations into Equation \eqref{eq:lambda}, the moving-lens signal can be written as a projected field,
  \begin{equation} \label{eq:W_lambda}
      \lambda(\nv) = \int\mathrm{d}\chi\, W_\lambda(\chi)\, \Lambda(\chi\nv), \quad\quad W_\lambda(\chi) \equiv -3 H_0^2 \Omega_{\rm m} Hf,
  \end{equation}
  with three-dimensional counterpart
  \begin{equation} \label{eq:Lambda}
      \Lambda(\mathbf{k}) = \int_{\mathbf{k}} \mathrm{D}k_1 \, \mathrm{D}k_2 \, \frac{\mathbf{k}_1 \cdot \mathbf{k}_2}{k_1^2 k_2^2} \, \delta_v(\mathbf{k}_1) \, \delta_{\rm m}(\mathbf{k}_2),
  \end{equation}
  where we have defined
  \begin{equation}
      \int_{\mathbf{k}} \mathrm{D}k_1 \dots \mathrm{D}k_n \equiv \int \frac{\mathrm{d}^3 k_1}{(2\pi)^3} \dots \frac{\mathrm{d}^3 k_n}{(2\pi)^3} \, (2\pi)^3 \, \delta^{\mathcal{D}}(\mathbf{k}_1 + \dots + \mathbf{k}_n - \mathbf{k}).
  \end{equation}
  The moving lens temperature is thus a quadratic functional of the matter field, with a kernel that is symmetric under $\mathbf{k}_1 \leftrightarrow \mathbf{k}_2$ and that depends on the relative orientation of the two momenta only through $\mathbf{k}_1\cdot\mathbf{k}_2$. 
  
  It is important to note that the observable given by Equation~\eqref{eq:lambda} is the full scalar product $v_i \partial_i \phi$, whereas the moving lens effect is sourced only by the components of the velocity and the potential gradient transverse to the line of sight. The two expressions differ by the radial contribution $v_\parallel \nabla_\parallel\phi$, which we show in Appendix~\ref{app:radial} to vanish in the squeezed limit and to be suppressed by $\mathcal{O}(q^2/k^2)$ beyond it. This is a stronger suppression than the residual foreground contamination derived in Section~\ref{ssec:foregrounds.beyond_squeezed}, which first appears at $\mathcal{O}(q/k)$, so the radial term is negligible wherever the foreground treatment is valid. We therefore neglect this difference and retain the full scalar product throughout the remainder of the analysis.
  
 \subsection{Projected galaxy momentum} \label{ssec:ml.galaxy_momentum}
  We define the galaxy momentum field as
  \begin{equation}
      \mathbf{q}_{\rm g}(\mathbf{x}) \equiv \delta_{\rm g}(\mathbf{x})\, \mathbf{v}_{\rm g}(\mathbf{x}),
  \end{equation}
  where $\delta_{\rm g}$ is the galaxy overdensity and $\mathbf{v}_{\rm g}$ is the reconstructed large-scale velocity field,
  \begin{equation}
      \mathbf{v}_{\rm g}(\mathbf{k}) \equiv \frac{i\mathbf{k}}{k^2} aHf\, \Theta(k)\, \delta_{\rm g}(\mathbf{k}).
  \end{equation}
  Here, $\Theta(k)$ is the low-pass filter that retains only modes with $k<k_L$. In Fourier space,
  \begin{equation}
      \mathbf{q}_{\rm g}(\mathbf{k}) = iaHf \int_{\mathbf{k}}\mathrm{D}k_1\,\mathrm{D}k_2\, \frac{\mathbf{k}_1}{k_1^2}\, \Theta(k_1)\, \delta_{\rm g}(\mathbf{k}_1)\, \delta_{\rm g}(\mathbf{k}_2).
  \end{equation}
  The field used to measure the ML effect is the projected transverse momentum,
   \begin{equation}
      \boldsymbol{\pi}_{\rm g}(\nv) = \int\mathrm{d}\chi\, W_{\pi}(\chi)\, (\mathbb{I}-\nv\nv^\dagger)\, \mathbf{q}_{\rm g}(\chi\nv), \quad\quad W_{\pi}(\chi) = H(z)\, p(z),
   \end{equation}
   where $p(z)$ is the redshift distribution of the galaxy sample and the projector $\mathbb{I}-\nv\nv^\dagger$ retains only the components tangent to the sky.

   Before the projection, it is useful to split the three-dimensional momentum into pieces longitudinal and transverse to its Fourier wavevector (i.e. scalar and vector modes, respectively),
   \begin{equation} \label{eq:perp_par}
       \mathbf{q}_{\rm g}(\mathbf{k}) = \mathbf{q}_{\rm g}^{\rm L}(\mathbf{k}) + \mathbf{q}_{\rm g}^{\rm T}(\mathbf{k}), \quad\quad \mathbf{q}_{\rm g}^{\rm L} = (\hat{\mathbf{k}}\cdot\mathbf{q}_{\rm g})\hat{\mathbf{k}}, \quad\quad \hat{\mathbf{k}}\cdot\mathbf{q}_{\rm g}^{\rm T} = 0.
   \end{equation}
   Throughout, the superscripts ${\rm L}$ and ${\rm T}$ denote components longitudinal and transverse to the Fourier wavevector $\hat{\mathbf{k}}$, whereas subscripts $\parallel$ and $\perp$ denote components parallel and perpendicular to the line of sight. The two decompositions are distinct: the one with respect to the Fourier wavevector is performed on the three-dimensional field before the line-of-sight projection, and is carried out to be consistent with the curved-sky calculation of Section~\ref{sec:full_sky}. Keeping the two calculations in the same basis allows the flat- and full-sky results to be compared term by term.

   The projected momentum can similarly be written as
   \begin{equation}
       \boldsymbol{\pi}_{\rm g}(\nv) = \boldsymbol{\pi}_{\rm g}^{\rm L}(\nv) + \boldsymbol{\pi}_{\rm g}^{\rm T}(\nv),
   \end{equation}
   with
   \begin{equation}
       \boldsymbol{\pi}_{\rm g}^{\rm L}(\nv) = \int\mathrm{d}\chi\, W_\pi(\chi) \int\frac{\mathrm{d}^3 k}{(2\pi)^3}\, e^{i\mathbf{k}\cdot(\chi\nv)}\, q_{\rm g}^{\rm L}(\mathbf{k})\, (\mathbb{I} - \nv\nv^\dagger)\, \hat{\mathbf{k}},
   \end{equation}
   \begin{equation}
       \boldsymbol{\pi}_{\rm g}^{\rm T}(\nv) = \int\mathrm{d}\chi\, W_\pi(\chi) \int\frac{\mathrm{d}^3 k}{(2\pi)^3}\, e^{i\mathbf{k}\cdot(\chi\nv)}\, (\mathbb{I} - \nv\nv^\dagger)\, \mathbf{q}_{\rm g}^{\rm T}(\mathbf{k}).
   \end{equation}
   For the longitudinal piece, the projection gives simply,
   \begin{equation}
       (\mathbb{I}-\nv\nv^\dagger)(q_{\rm g}^{\rm L} \hat{\mathbf{k}}) = q_{\rm g}^{\rm L}(\hat{\mathbf{k}}-x\nv), \quad\quad x \equiv \hat{\mathbf{k}}\cdot\nv.
   \end{equation}

   We now specialise to a sufficiently small patch on the sky, choosing coordinates such that the line of sight is aligned with the $z$-axis, $\nv \simeq \hat{\mathbf{z}}$, and neglecting the variation of the line-of-sight direction across the patch. Writing $\hat{\mathbf{k}} = (\mathbf{k}_\perp + k_\parallel \hat{\mathbf{z}})/k$, we have
   \begin{equation}
       \hat{\mathbf{k}}-x\nv \simeq \frac{\mathbf{k}_\perp}{k} = \frac{k_\perp}{k} \hat{\mathbf{k}}_\perp.
   \end{equation}
   Performing the angular integral,
   \begin{equation}
       \int\mathrm{d}^2\theta\, e^{-i\mathbf{l}\cdot\boldsymbol{\theta}}\,e^{i\chi\mathbf{k}_\perp\cdot\boldsymbol{\theta}} = (2\pi)^2 \delta^{\mathcal{D}}(\mathbf{l}-\chi\mathbf{k}_\perp),
   \end{equation}
   the 2D Fourier-space momentum components become
   \begin{equation}
       \boldsymbol{\pi}^{\rm L}(\mathbf{l}) = \int\frac{\mathrm{d}\chi}{\chi^2}\, W_\pi(\chi) \int\frac{\mathrm{d}k_\parallel}{2\pi}\, e^{i\chi k_\parallel}\, \left.q_{\rm g}^{\rm L}(\mathbf{k})\, \frac{k_\perp}{k}\, \hat{\mathbf{k}}_\perp\right|_{\mathbf{k}_\perp=\mathbf{l}/\chi},
   \end{equation}
   \begin{equation}
       \boldsymbol{\pi}^{\rm T}(\mathbf{l}) = \int\frac{\mathrm{d}\chi}{\chi^2}\, W_\pi(\chi) \int\frac{\mathrm{d}k_\parallel}{2\pi}\, e^{i\chi k_\parallel} \left.(\mathbb{I}-\hat{\mathbf{z}}\hat{\mathbf{z}}^\dagger)\, \mathbf{q}_{\rm g}^{\rm T}(\mathbf{k})\right|_{\mathbf{k}_\perp=\mathbf{l}/\chi}.
   \end{equation}
   For a spin-1 field on the flat sky, the $E$- and $B$-modes are given by
   \begin{equation}
       \pi_E(\mathbf{l}) = -i \hat{\mathbf{l}} \cdot \boldsymbol{\pi}(\mathbf{l}), \quad\quad \pi_B(\mathbf{l}) = -i\hat{\mathbf{l}} \times \boldsymbol{\pi}({\mathbf{l}}).
   \end{equation}
   Since $\hat{\mathbf{l}} \parallel \hat{\mathbf{k}}_\perp$, the longitudinal component is purely a gradient,
   \begin{equation} \label{eq:pi_E_par}
       \pi_E^{\rm L}(\mathbf{l}) = -i \int\frac{\mathrm{d}\chi}{\chi^2}\, W_\pi(\chi) \int\frac{\mathrm{d}k_\parallel}{2\pi}\, e^{ik_\parallel\chi}\, q_{\rm g}^{\rm L}(\mathbf{k}) \frac{k_\perp}{k}, \quad\quad \pi_B^{\rm L} = 0,
   \end{equation}
   while the transverse component sources both $E$- and $B$-modes,
   \begin{equation} \label{eq:pi_E_perp}
       \pi_E^{\rm T}(\mathbf{l}) = -i\int\frac{\mathrm{d}\chi}{\chi^2}\, W_\pi(\chi) \int \frac{\mathrm{d}k_\parallel}{2\pi}\, e^{ik_\parallel\chi}\, \hat{\mathbf{l}}\cdot(\mathbb{I}-\hat{\mathbf{z}}\hat{\mathbf{z}}^\dagger)\,\mathbf{q}_{\rm g}^{\rm T}(\mathbf{k}),
   \end{equation}
   \begin{equation} \label{eq:pi_B_perp}
       \pi_B^{\rm T}(\mathbf{l}) = -i\int\frac{\mathrm{d}\chi}{\chi^2}\, W_\pi(\chi) \int\frac{\mathrm{d}k_\parallel}{2\pi}\, e^{ik_\parallel\chi}\, \hat{\mathbf{l}}\times(\mathbb{I}-\hat{\mathbf{z}}\hat{\mathbf{z}}^\dagger)\,\mathbf{q}_{\rm g}^{\rm T}(\mathbf{k}).
   \end{equation}
   The fact that $\pi_B^{\rm L}=0$ identically is immediate since the longitudinal momentum points along $\hat{\mathbf{k}}$, whose sky projection is parallel to $\hat{\mathbf{l}}$, and the cross product of parallel vectors vanishes. This is the flat-sky counterpart of the curved-sky result of Section~\ref{sec:full_sky}.

   The projected $E$- and $B$-modes derived above can be written more compactly by introducing three-dimensional counterparts, in the same way that $\Lambda(\mathbf{k})$ in Equation~\eqref{eq:Lambda} is the three-dimensional source of the projected temperature $\lambda(\nv)$. We therefore define $\Pi_E(\mathbf{k})$ and $\Pi_B(\mathbf{k})$ through
   \begin{equation}
       \pi_{E,B}(\mathbf{l}) = \int\frac{\mathrm{d}\chi}{\chi^2}\, \tilde{W}_\pi(\chi) \int\frac{\mathrm{d}k_\parallel}{2\pi}\, e^{ik_\parallel \chi} \left.\Pi_{E,B}(\mathbf{k})\right|_{\mathbf{k}_\perp = \mathbf{l}/\chi}, \quad\quad \tilde{W}_\pi(\chi) \equiv aHf\, W_\pi(\chi),
   \end{equation}
   where the factor $aHf$ carried by the reconstructed velocity has been absorbed into the radial kernel. Comparing with Equations~\eqref{eq:pi_E_par}--\eqref{eq:pi_B_perp} and using $\mathbf{q}_{\rm g}(\mathbf{k}) \propto i\mathbf{k}_1/k_1^2$, the explicit factor of $i$ cancels against the $-i$ in the definition of the $E$- and $B$-modes, leaving
   \begin{equation} \label{eq:Pi_E}
       \Pi_E(\mathbf{k}) = \int_{\mathbf{k}}\mathrm{D}k_1\,\mathrm{D}k_2\, \frac{\mathbf{k}_1 \cdot \mathbf{k}_\perp}{k_1^2 k_\perp}\, \Theta(k_1)\, \delta_{\rm g}(\mathbf{k}_1)\, \delta_{\rm g}(\mathbf{k}_2),
   \end{equation}
   \begin{equation} \label{eq:Pi_B}
       \Pi_B(\mathbf{k}) = \int_{\mathbf{k}}\mathrm{D}k_1\,\mathrm{D}k_2\, \frac{\mathbf{k}_1 \times \mathbf{k}_\perp}{k_1^2 k_\perp}\, \Theta(k_1)\, \delta_{\rm g}(\mathbf{k}_1)\, \delta_{\rm g}(\mathbf{k}_2).
   \end{equation}
   The $E$ and $B$ labels are inherited from the two-dimensional decomposition: $\Pi_{E,B}$ are three-dimensional fields, but defined relative to a fixed line of sight, entering through $\mathbf{k}_\perp$. Splitting these according to Equation~\eqref{eq:perp_par}, the longitudinal $E$-mode kernel is
   \begin{equation} \label{eq:Pi_E_par}
       \Pi_E^{\rm L}(\mathbf{k}) = \int_{\mathbf{k}}\mathrm{D}k_1\,\mathrm{D}k_2\, \frac{k_\perp(\mathbf{k} \cdot \mathbf{k}_1)}{k^2 k_1^2}\, \Theta(k_1)\, \delta_{\rm g}(\mathbf{k}_1)\, \delta_{\rm g}(\mathbf{k}_2),
   \end{equation}
   and, from $\Pi_E^{\rm T} = \Pi_E - \Pi_E^{\rm L}$
   \begin{equation} \label{eq:Pi_E_perp}
       \Pi_E^{\rm T}(\mathbf{k}) = \int_{\mathbf{k}}\mathrm{D}k_1\,\mathrm{D}k_2\, \left[\frac{\mathbf{k}_1 \cdot \mathbf{k}_\perp}{k_1^2 k_\perp} - \frac{k_\perp(\mathbf{k} \cdot \mathbf{k}_1)}{k^2 k_1^2}\right] \Theta(k_1)\, \delta_{\rm g}(\mathbf{k}_1)\, \delta_{\rm g}(\mathbf{k}_2),
   \end{equation}
   with $\Pi_B^{\rm L} = 0$ and $\Pi_B^{\rm T} = \Pi_B$.

   In the Limber limit, $k_\parallel \ll k_\perp$ so $k \simeq k_\perp$ and Equations~\eqref{eq:Pi_E_par}-\eqref{eq:Pi_E_perp} collapse to
   \begin{equation}
       \Pi_E^{\rm L}(\mathbf{k}) \simeq \Pi_E(\mathbf{k}), \quad\quad \Pi_E^{\rm T}(\mathbf{k}) \simeq 0.
   \end{equation}
   The flat-sky $E$-mode is therefore recovered entirely from the momentum component longitudinal to $\mathbf{k}$, while the $B$-mode is sourced only by the transverse component. This is the flat-sky counterpart of the result we derive on the full-sky in Section~\ref{sec:full_sky}.

 \subsection{The moving lens power spectrum} \label{ssec:ml.power_spectrum}
  Under the flat-sky and Limber approximations, the angular power spectrum of two projected fields is related to the three-dimensional power spectrum of their sources by
  \begin{equation}
      C_\ell^{uv} = \int \frac{\mathrm{d}\chi}{\chi^2}\, W_u(\chi)\, W_v(\chi)\, P_{UV}(k_\parallel=0, k_\perp=\ell/\chi).
  \end{equation}
  For the cross-correlation of the moving lens temperature anisotropy with the momentum $E$-mode,
   \begin{equation} \label{eq:Cl_pi_lambda_flat}
       C_\ell^{\pi_E\lambda} = \int \frac{\mathrm{d}\chi}{\chi^2} \, \tilde{W}_{\pi}(\chi) \, W_{\lambda}(\chi) \, P_{\Pi_E \Lambda}(0,\ell/\chi),
   \end{equation}
   where $P_{\Pi_E \Lambda}$ follows from the expectation value of the product of Equations \eqref{eq:Lambda} and \eqref{eq:Pi_E}:
   \begin{equation}
       \left\langle \Pi_E(\mathbf{k}) \Lambda(\mathbf{k}') \right\rangle = \int_{\mathbf{k}} \mathrm{D}k_1 \, \mathrm{D}k_2 \, \Theta(k_1) \, \frac{\mathbf{k}_1 \cdot \mathbf{k}_\perp}{k_1^2 k_\perp} \int_{\mathbf{k}'} \mathrm{D}q_1 \, \mathrm{D}q_2 \, \frac{\mathbf{q}_1 \cdot \mathbf{q}_2}{q_1^2 q_2^2} \left\langle \delta_{\rm g}(\mathbf{k}_1) \delta_{\rm g}(\mathbf{k}_2) \delta_v(\mathbf{q}_1) \delta_{\rm m}(\mathbf{q}_2) \right\rangle .
   \end{equation}
   The four-point function splits into two non-vanishing disconnected pieces, given by Wick's theorem, and a connected piece. We will express these in terms of the power spectra of the different fields, and their connected trispectrum, defined as:
   \begin{equation}
       \langle \delta_a(\mathbf{k}_1) \delta_b(\mathbf{k}_2) \rangle \equiv (2\pi)^3 \delta^{\mathcal{D}}(\mathbf{k}_1+\mathbf{k}_2)\, P_{ab}(k_1),
   \end{equation}
   \begin{equation} \label{eq:four_point}
       \langle \delta_a(\mathbf{k}_1) \delta_b(\mathbf{k}_2) \delta_c(\mathbf{k}_3) \delta_d(\mathbf{k}_4) \rangle_{\rm c} \equiv (2\pi)^3 \delta^{\mathcal{D}}(\mathbf{k}_1+\mathbf{k}_2+\mathbf{k}_3+\mathbf{k}_4)\, \mathcal{T}_{abcd}(\mathbf{k}_1, \mathbf{k}_2, \mathbf{k}_3, \mathbf{k}_4).
   \end{equation}

   \subsubsection{Disconnected contribution} \label{sssec:ml.power_spectrum_disconnected}
    At leading order, the four-point function reduces to two Wick contractions, which are given by $\Theta(k_1)P_{{\rm g}v}(k_1)P_{\rm gm}(k_2)$ and $\Theta(k_1)P_{\rm gm}(k_1)P_{{\rm g}v}(k_2)$. On large scales, $\delta_v\simeq \delta_{\rm m}$, and both contractions are identical, yielding an overall factor of two. As discussed earlier, however, in our simplified model the velocity field entering the ML signal is sourced by the linear matter overdensity, and lacks the usual small-scale ``1-halo'' enhancement. This breaks the symmetry between both terms, with the first one dominating on small scales, as we will see.

    The first contraction, in which the first galaxy field pairs with the velocity and the second with the matter field, gives
    \begin{equation}
         P_{\Pi_E \Lambda}^{\rm (dc,1)}(k_\perp) = \int\frac{\mathrm{d}^3 q}{(2\pi)^3}\, \frac{\mathbf{q}\cdot(\mathbf{k}-\mathbf{q})}{q^2 |\mathbf{k}-\mathbf{q}|^2} \, \frac{\mathbf{q}\cdot\mathbf{k}_\perp}{q^2 k_\perp}\, \Theta(q)\, P_{\mathrm{g}v}(q)\, P_{\mathrm{gm}}(|\mathbf{k}-\mathbf{q}|).
    \end{equation}
    In the squeezed limit $k_\perp\gg k_L>q$,
    \begin{equation}
        P_{\Pi_E \Lambda}^{\rm (dc,1)}(k_\perp)  \simeq \frac{P_{\rm gm}(k_\perp)}{k_\perp} \int\frac{\mathrm{d}^3 q}{(2\pi)^3}\, \frac{(\mathbf{q} \cdot \hat{\mathbf{k}}_\perp)^2}{q^4}\, \Theta(q) \, P_{{\rm g}v}(q) = \frac{P_{\rm gm}(k_\perp)}{k_\perp}\int_0^{k_L}\frac{\mathrm{d}q}{6\pi^2}\, P_{{\rm g}v}(q).
    \end{equation}
    The angular integrand is quadratic in $\mu \equiv \hat{\mathbf{q}} \cdot \hat{\mathbf{k}}_\perp$ and so has a non-zero angular average, $\langle \mu^2 \rangle = 1/3$. The filter thus does not remove the signal; it only truncates the $q$-integral, which is an overall amplitude.

    The second contraction gives the same expression with $P_{\mathrm{g}v} \leftrightarrow P_{\mathrm{gm}}$,
    \begin{equation}
        P_{\Pi_E \Lambda}^{\rm (dc,2)}(k_\perp\gg k_L) = \frac{P_{\mathrm{g}v}(k_\perp)}{ k_\perp}\int_0^{k_L}\frac{\mathrm{d}q}{6\pi^2}\, P_{\mathrm{gm}}(q),
    \end{equation}
    so that
    \begin{equation} \label{eq:P_dc_total}
        P_{\Pi_E \Lambda}^{\rm (dc)}(k_\perp\gg k_L) = \frac{1}{3k_\perp}\left[P_{\rm gm}(k_\perp)\int_0^{k_L}\frac{\mathrm{d}q}{2\pi^2}\, P_{{\rm g}v}(q) + P_{{\rm g}v}(k_\perp)\int_0^{k_L}\frac{\mathrm{d}q}{2\pi^2}\, P_{\rm gm}(q)\right].
    \end{equation}
    Although formally symmetric, the two terms behave very differently. In $P^{\rm(dc,1)}$, the small-scale behaviour is governed by $P_{\mathrm{gm}}(k_\perp)$, which is enhanced by non-linear growth through its one-halo term. In $P^{\rm (dc,2)}$, on the other hand, the small-scale factor is $P_{\mathrm{g}v}(k_\perp)$, which lacks this enhancement, and is therefore suppressed by comparison (by up to 2-3 orders of magnitude at $z\sim0.55$). Hence, at large $k_\perp$,
    \begin{equation} \label{eq:ml_signal}
        P_{\Pi_E \Lambda}^{\rm (dc)}(k_\perp\gg k_L) \simeq \frac{P_{\rm gm}(k_\perp)}{3k_\perp}\int_0^{k_L}\frac{\mathrm{d}q}{2\pi^2}\, P_{{\rm g}v}(q).
    \end{equation}
    In turn, on large scales, where non-linear effects are negligible, the two contractions become equal, as we show numerically in Section~\ref{ssec:results.ml_signal}.

    It is convenient to label the two filtered integrals appearing in Equation~\eqref{eq:P_dc_total},
    \begin{equation} \label{eq:A_def}
        \mathcal{A}_X(k_L) \equiv \int_0^{k_L}\frac{\mathrm{d}q}{2\pi^2}\, P_{\mathrm{g}X}(q), \quad\quad X=\{v,\,{\rm m}\},
    \end{equation}
    since they are the only route by which the reconstructed velocity cut enters the moving-lens signal on small scales. These amplitudes are velocity dispersions in all but normalisation, with the factors of $aHf$ having been absorbed into the projection kernels: $\langle \mathbf{v}_{\rm g}\cdot\mathbf{v}\rangle = (aHf)^2\,\mathcal{A}_v(k_L)$.
   
   \subsubsection{Connected contribution} \label{sssec:ml.power_spectrum_connected}
    Using Equation~\eqref{eq:four_point}, the connected piece can be written as
    \begin{equation}
        P_{\Pi_E \Lambda}^{\rm (c)}(k_\perp) = \int\frac{\mathrm{d}^3 p}{(2\pi)^3} \int\frac{\mathrm{d}^3 q}{(2\pi)^3}\, \frac{\mathbf{p}\cdot\mathbf{k}_\perp}{p^2 k_\perp}\, \Theta(p)\, \frac{-\mathbf{q}\cdot(\mathbf{k}+\mathbf{q})}{q^2 |\mathbf{k}+\mathbf{q}|^2}\, \mathcal{T}_{{\rm gg}v{\rm m}}(\mathbf{p}, \mathbf{k}-\mathbf{p}, \mathbf{q}, -\mathbf{k}-\mathbf{q}),
    \end{equation}
    which may be evaluated, for example, within the framework of the halo model.

  \subsubsection[Vanishing of the $B$-mode correlation]{\boldmath Vanishing of the $B$-mode correlation} \label{sssec:ml.power_spectrum_bmode}
   We now repeat the calculation for the $B$-mode. From Equation~\eqref{eq:Pi_B}, the correlator $\langle \Pi_B(\mathbf{k}) \Lambda(\mathbf{k}') \rangle$ has exactly the same structure as its $E$-mode counterpart, with the single replacement $\mathbf{k}_1 \cdot \mathbf{k}_\perp \to \mathbf{k}_1 \times \mathbf{k}_\perp$, where the cross product of two vectors tangent to the sky is understood as a pseudoscalar obtained by projecting along $\nv$. Every step of Section~\ref{sssec:ml.power_spectrum_disconnected} therefore carries over unchanged, including the inequivalence of the two Wick contractions. The first contraction gives
   \begin{equation} \label{eq:P_B_1}
       P_{\Pi_B \Lambda}^{\rm (dc,1)}(k_\perp) = \int\frac{\mathrm{d}^3 q}{(2\pi)^3}\, \frac{\mathbf{q}\cdot(\mathbf{k}-\mathbf{q})}{q^2 |\mathbf{k}-\mathbf{q}|^2} \, \frac{\mathbf{q}\times\mathbf{k}_\perp}{q^2 k_\perp}\, \Theta(q)\, P_{\mathrm{g}v}(q)\, P_{\mathrm{gm}}(|\mathbf{k}-\mathbf{q}|),
   \end{equation}
   and the second the same expression with $P_{\mathrm{g}v} \leftrightarrow P_{\mathrm{gm}}$.

   We decompose the soft mode as $\mathbf{q}=\mathbf{q}_\perp + q_\parallel \nv$ and let $\varphi$ denote the azimuthal angle of $\mathbf{q}_\perp$ about the line of sight, measured from $\hat{\mathbf{k}}_\perp$. The component of $\mathbf{q}$ along $\nv$ thus does not contribute to the cross product, since $\nv \cdot (\nv \times \hat{\mathbf{k}}_\perp) = 0$, so
   \begin{equation}
       \mathbf{q} \times \hat{\mathbf{k}}_\perp = q_\perp\sin\varphi, \quad\quad \mathbf{q} \cdot \hat{\mathbf{k}}_\perp = q_\perp\cos\varphi.
   \end{equation}
   The $B$-mode kernel is therefore odd under the reflection $\varphi \to -\varphi$. Every remaining factor in Equation~\eqref{eq:P_B_1} is even under the same reflection. In the Limber limit, in which $\mathbf{k} = \mathbf{k}_\perp$, we therefore have 
       \begin{equation}
        \mathbf{q}\cdot(\mathbf{k}-\mathbf{q}) = q_\perp k_\perp \cos\varphi - q^2, \quad\quad
        |\mathbf{k}-\mathbf{q}|^2 = k_\perp^2 + q^2 - 2 q_\perp k_\perp \cos\varphi,
    \end{equation}
    both of which depend on $\varphi$ only through $\cos\varphi$, while the two power spectra depend only on the magnitudes $q$ and $|\mathbf{k}-\mathbf{q}|$. The azimuthal integrand is thus odd and each contraction vanishes separately, 
    \begin{equation}
        P_{\Pi_B \Lambda}^{\rm (dc)}(k_\perp) = 0.
    \end{equation}
    Hence, the result is completely insensitive to the asymmetry between $P_{\mathrm{g}v}$ and $P_{\mathrm{gm}}$ that controls the amplitude of the $E$-mode signal, and to the modelling of the power spectra altogether. Furthermore, the cancellation does not require the squeezed limit and instead follows from the reflection symmetry alone.

    We emphasise that this is a statement about the cross-correlation, not about the field itself. The momentum $B$-mode is non-zero at the map level: what vanishes is its correlation with a scalar field on the sky, as parity requires. Section~\ref{ssec:full_sky.spin-1} establishes the same conclusion on the full sky, without invoking either the flat-sky or Limber approximation.

\section{Foreground contamination} \label{sec:foregrounds}
 Up to this point, we have described the moving-lens signal. In practice, however, under imperfect component separation, any CMB temperature map also contains residual astrophysical foreground emission. Since these foregrounds trace the same underlying large-scale structure used to reconstruct the galaxy momentum field, they generate a non-zero contribution to the moving-lens estimator \cite{2305.15462}. In this section, we derive the form of this contamination before investigating its behaviour in and beyond the squeezed limit.

 \subsection{The foreground power spectrum} \label{ssec:foregrounds.power_spectrum}
  We model the foreground contribution as a projected tracer of an underlying statistically isotropic three-dimensional field $\delta_F$,
  \begin{equation} \label{eq:delta_f}
      \delta_f(\nv) = \int\mathrm{d}\chi\, W_{f}(\chi)\, \delta_F(\chi\nv),
  \end{equation}
  where $W_{f}(\chi)$ denotes the radial kernel of the foreground field. The most relevant examples of such foregrounds are the Cosmic Infrared Background, where $\delta_F$ traces the fluctuations in the infrared emissivity, or the star formation rate density \cite{1801.10146}, and the thermal Sunyaev--Zel'dovich effect, where $\delta_F$ traces fluctuations in the electron pressure \cite{astro-ph/0205468}. In both cases, $\delta_F$ is strongly correlated with the matter overdensity $\delta_{\rm m}$, and is a biased tracer of it.
  
  The cross-spectrum of $\delta_f$ with the momentum $E$-mode is
  \begin{equation}
      C_\ell^{\pi_E f} = \int\frac{\mathrm{d}\chi}{\chi^2}\, \tilde{W}_{\pi}(\chi)\, W_{f}(\chi)\, P_{\Pi_E F}(\ell/\chi),
  \end{equation}
  with
  \begin{equation}
      \left\langle \Pi_E(\mathbf{k}) \delta_F(\mathbf{k}') \right\rangle = \int_{\mathbf{k}} \mathrm{D}k_1\, \mathrm{D}k_2\, \frac{\mathbf{k}_1\cdot\mathbf{k}_\perp}{k_1^2 k_\perp}\, \Theta(k_1)\, \left\langle \delta_{\rm g}(\mathbf{k}_1) \delta_{\rm g}(\mathbf{k}_2) \delta_F(\mathbf{k}') \right\rangle.
  \end{equation}
  The three-point correlation function can be expressed in terms of the galaxy--galaxy--foreground bispectrum,
  \begin{equation}
      \left\langle \delta_{\rm g}(\mathbf{k}_1) \delta_{\rm g}(\mathbf{k}_2) \delta_F(\mathbf{k}') \right\rangle = (2\pi)^3 \delta^{\mathcal{D}}(\mathbf{k}_1 + \mathbf{k}_2 + \mathbf{k}') \, B_{\mathrm{gg}F}(\mathbf{k}_1, \mathbf{k}_2, \mathbf{k}').
  \end{equation}
  The two delta functions together enforce $\mathbf{k}'=-\mathbf{k}$, and relabelling $\mathbf{q}\equiv\mathbf{k}_1$ gives
  \begin{equation} \label{eq:foreground_pk}
      P_{\Pi_E F}(k_\perp) = \int \frac{\mathrm{d}^3 q}{(2\pi)^3} \, \frac{\mathbf{q} \cdot \hat{\mathbf{k}}_\perp}{q^2} \, \Theta(q) \, B_{\mathrm{gg}F}(\mathbf{q}, \mathbf{k}-\mathbf{q}, -\mathbf{k}).
  \end{equation}
  This expression demonstrates that the foreground contamination is entirely determined by the galaxy--galaxy--foreground bispectrum. The behaviour of this bispectrum in the squeezed limit therefore governs the response of foreground contamination to the low-pass filtering used in velocity reconstruction, represented by $\Theta(q)$ above. We first consider the squeezed limit, in which the reconstructed velocity mode is much longer than the transverse mode contributing to the foreground signal, $q<k_L \ll k_\perp$.

 \subsection{The squeezed limit} \label{ssec:foregrounds.squeezed_limit}
  In the squeezed configuration $q \ll k_\perp$, the bispectrum can be expressed in the so-called response formalism \cite{1503.03487,1701.03374,1703.09212,2212.11940}:
  \begin{equation} \label{eq:response}
      B_{\mathrm{gg}F}(\mathbf{q}, \mathbf{k}_\perp-\mathbf{q}, -\mathbf{k}_\perp) \simeq P_{F\mathrm{g}}(k_\perp)\,{\cal R}_{F\mathrm{g}}(k_\perp)\,P_{\rm gm}(q),
  \end{equation}
  which states that the effect of a long-wavelength mode on the short-scale power spectrum is captured entirely by the response $\mathcal{R}_{F\mathrm{g}}$ of that power spectrum to a change in the local background density:
  \begin{equation}
    {\cal R}_{ab}(k)=\left.\frac{\partial \log P_{ab}(k)}{\partial \delta_{\rm m}({\bf q})}\right|_{q\ll k}.
  \end{equation}
  At leading order, the response depends only on the amplitude of the long mode, not on its direction. Hence, in the squeezed limit
  \begin{equation}
      P_{\Pi_E F}(k_\perp\gg k_L) \simeq P_{F\mathrm{g}}(k_\perp)\, \mathcal{R}_{F\mathrm{g}}(k_\perp) \int_0^{k_L}\frac{\mathrm{d}q\,q}{(2\pi)^2}P_{\rm gm}(q)\int_{-1}^1 \mathrm{d}\mu\,\mu= 0,
  \end{equation}
  the angular integral being odd in $\mu\equiv\hat{\bf q}\cdot\hat{\bf k}_\perp$. 
  
  Equation~\eqref{eq:response} retains only the isotropic part of the response. More generally, the short-scale power spectrum also responds to the long-wavelength tidal field \cite{Halder2022}, adding a term proportional to $\mathcal{R}^{K}_{F\mathrm{g}}(k_\perp)\, P_{\rm gm}(q)\, P_2(\mu)$, where $\mathcal{R}^{K}_{F\mathrm{g}}$ is the response to the tidal field and $P_2$ is the second Legendre polynomial. This does not affect our conclusion: the tidal response enters through $\mu^2$ and is therefore even, so the angular integrand remains odd and the resulting integral vanishes.
  
  The cancellation holds for any foreground satisfying Equation~\eqref{eq:delta_f} and is independent of its astrophysical nature (as long as it traces the LSS), the specific foreground spectral energy distribution (SED), or details of the velocity reconstruction filter beyond it being isotropic. This is the analytical explanation of the empirical behaviour reported in \cite{Hotinli2026}: the low-$k$ filter forces the foreground bispectrum into precisely the configuration in which it decouples from the momentum estimator by symmetry. By contrast, the angular integrand of the signal given by Equation~\eqref{eq:ml_signal} is even in $\mu$. The signal is thus preserved in the squeezed configuration, whereas foreground contamination vanishes at leading order. The residual contamination from foregrounds therefore arises only from corrections to the squeezed-limit expansion, which we consider in the following section.

 \subsection{Beyond the squeezed limit} \label{ssec:foregrounds.beyond_squeezed}
  In the previous section, we showed that the foreground contribution vanishes identically in the squeezed limit due to angular symmetry. In practice, however, the reconstructed velocity modes are finite and the ratio $\epsilon \equiv q/k$ is not strictly zero. The residual foreground contamination is therefore governed by the leading corrections in an expansion about the squeezed limit. 
  
  We adopt a halo model framework to derive this residual contamination, writing the halo-model bispectrum \cite{Valageas2011} as
  \begin{equation}
       B_{\mathrm{gg}F} = B_{\mathrm{gg}F}^{\rm 1h} + B_{\mathrm{gg}F}^{\rm 2h} + B_{\mathrm{gg}F}^{\rm 3h},
  \end{equation}
  where
  \begin{equation} \label{eq:B_1h_def}
      B_{\delta_A \delta_B \delta_C}^{\rm 1h}(\mathbf{k}_1, \mathbf{k}_2, \mathbf{k}_3) = \int\mathrm{d}M\, n(M)\, \left\langle u_{\delta_A}(k_1|M) u_{\delta_B}(k_2|M) u_{\delta_C}(k_3|M) \right\rangle,
  \end{equation}
  \begin{equation} \label{eq:B_2h_def}
      B_{\delta_A \delta_B \delta_C}^{\rm 2h}(\mathbf{k}_1, \mathbf{k}_2, \mathbf{k}_3) = I_{\delta_A \delta_B}^1(k_1, k_2) \, I_{\delta_C}^1(k_3) \, P_{\rm lin}(k_3) + \text{cyclic permutations,}
  \end{equation}
  \begin{equation} \label{eq:B_3h_def}
      B_{\delta_A \delta_B \delta_C}^{\rm 3h}(\mathbf{k}_1, \mathbf{k}_2, \mathbf{k}_3) = \bU{A}(k_1) \bU{B}(k_2) \bU{C}(k_3) \, B^{\rm PT}(\mathbf{k}_1, \mathbf{k}_2, \mathbf{k}_3).
  \end{equation}
  Here, we have used the definitions
   \begin{equation}
       \bU{A}(k) \equiv I_{\delta_A}^1(k) \equiv \int\mathrm{d}M\, n(M)\, b_h(M)\, \langle u_{\delta_A}(k|M) \rangle,
   \end{equation}
   \begin{equation}
       I_{\delta_A \delta_B}^1(k_1, k_2) \equiv \int \mathrm{d}M \, n(M) \, b_h(M) \, \langle u_{\delta_A}(k_1|M) u_{\delta_B}(k_2|M) \rangle.
   \end{equation}
  In every case, the mechanism is the same: expanding the short-wavelength legs about $k$ generates a term proportional to $q\mu$, which combines with the explicit $\mu$ of Equation~\eqref{eq:foreground_pk} to give a surviving $\mu^2$ contribution suppressed by one power of $\epsilon$. In the following subsections, we quote the results, relegating the detailed derivation to Appendix~\ref{app:beyond_squeezed}. We stress that this should be interpreted as a halo model estimate of the leading beyond-squeezed correction, rather than a complete squeezed limit expansion of the full bispectrum (see e.g. \cite{Chiang2014, Chiang2017, 2212.11940}).

  \subsubsection{One-halo contribution} \label{sssec:foregrounds.beyond_squeezed.1h}
   Expanding the short-wavelength argument as $|\mathbf{k}-\mathbf{q}| = k - q\mu + \mathcal{O}(q^2/k)$,
   \begin{equation}
       u_{\rm g}(|\mathbf{k}-\mathbf{q}|\,|M) = u_{\rm g}(k|M) - q\mu \frac{\partial u_{\rm g}(k|M)}{\partial k} + \mathcal{O}(q^2 \partial_k^2 u_{\rm g}).
   \end{equation}
   The zeroth-order term reproduces the squeezed-limit result and vanishes because the angular kernel is odd in $\mu$. The first non-zero contribution is then
   \begin{equation}
       P_{\Pi_E F}^{\rm 1h, (1)}(k_\perp) = -\frac{1}{6\pi^2} \int_0^{k_L} \mathrm{d}q \, q^2 \int \mathrm{d}M \, n(M) \, \left\langle u_{\rm g}(q|M) \, u_F(k|M) \, \frac{\partial u_{\rm g}(k|M)}{\partial k} \right\rangle. \label{eq:P_1h_1}
   \end{equation}
   The residual one-halo contamination is therefore controlled entirely by the scale dependence of the short-scale halo profile, $\partial_k u_g(k|M)$. Physically, the filter cancels the response of the small-scale power to a uniform change in the background density, and what survives is the sensitivity of the small-scale profile to the gradient of the long mode across the halo.
   
  \subsubsection{Two-halo contribution} \label{sssec:foregrounds.beyond_squeezed.2h}
   The two-halo term, given by Equation~\eqref{eq:B_2h_def}, has three contributions, each of which is expanded in Appendix~\ref{app:beyond_squeezed}. The zeroth-order term vanishes in each case and, summing the three pieces, the first non-zero correction is
   \begin{align}
       P_{\Pi_E F}^{\mathrm{2h},(1)}(k_\perp) = &-\frac{I_F^1(k) \, P_{\rm lin}(k)}{6\pi^2} \int_0^{k_L} \mathrm{d}q \, q^2 \, \partial_k I_{\rm gg}^1(q,k) \nonumber \\[0.2em]
       &- \frac{1}{6\pi^2} \left.\frac{\partial I_{\mathrm{g}F}^1(k_1,k_2)}{\partial k_1}\right|_{k_1=k_2=k} \int_0^{k_L} \mathrm{d}q \, q^2 \, I_{\rm g}^1(q) \, P_{\rm lin}(q) \nonumber \\[0.2em]
       &- \frac{P_{\rm lin}(k)}{6\pi^2} \int_0^{k_L} \mathrm{d}q \, q^2 \, I_{F\mathrm{g}}^1(k,q) \left[\frac{n_k}{k} I_{\mathrm{g}}^1(k) + \partial_k I_{\mathrm{g}}^1(k)\right], \label{eq:P_2h_1}
   \end{align}
   where $n_k \equiv \mathrm{d}\ln P_{\rm lin}(k)/\mathrm{d}\ln k$.
   
  \subsubsection{Three-halo contribution} \label{sssec:foregrounds.beyond_squeezed.3h}
   The three-halo term dominates on large scales and is the leading contribution to squeezed configurations since it contains the perturbation theory bispectrum,
   \begin{equation} \label{eq:B_PT_def}
       B^{\rm PT}(\mathbf{k}_1, \mathbf{k}_2, \mathbf{k}_3) = 2 F_2(\mathbf{k}_1, \mathbf{k}_2) \, P_{\rm lin}(k_1) \, P_{\rm lin}(k_2) + (2,3) + (3,1),
   \end{equation}
   which in turn depends on the linear matter power spectrum, $P_{\rm lin}(k)$, and the second-order density kernel, $F_2$, defined by
   \begin{equation} \label{eq:F2kernel}
       F_2(\mathbf{k}_1, \mathbf{k}_2) \equiv \frac{5}{7} + \frac{\mu_{12}}{2}\left(\frac{k_2}{k_1} + \frac{k_1}{k_2}\right) + \frac{2}{7}\mu_{12}^2, \quad\quad \mu_{12} \equiv \frac{\mathbf{k}_1 \cdot \mathbf{k}_2}{k_1 k_2}.
   \end{equation}
   Expanding the three $F_2$ kernels and $P_{\rm lin}(|\mathbf{k}-\mathbf{q}|)$ to second order in $\epsilon$ (Appendix~\ref{app:beyond_squeezed}), the tree-level bispectrum in our configuration takes the compact form
   \begin{equation} \label{eq:B_PT_config}
       B^{\rm PT}(\mathbf{q},\mathbf{k}-\mathbf{q},-\mathbf{k}) = P_{\rm lin}(q)\, P_{\rm lin}(k)\, \left[A_0 + A_2\mu^2 + \epsilon(A_1\mu + A_3\mu^3)\right],
   \end{equation}
   where the coefficients $\{A_i\}$ are given by
   \begin{equation} \label{eq:A_i}
       A_0 \equiv \frac{13}{7}, \quad A_2 \equiv \frac{8}{7}-n_k, \quad A_1 \equiv \frac{n_k-16}{14}, \quad A_3 \equiv \frac{8}{7} - \frac{11 n_k}{7} + \frac{n_k^2 + \alpha_k}{2},
   \end{equation}
   where $\alpha_k \equiv \mathrm{d}^2 \ln P_{\rm lin}(k)/\mathrm{d}(\ln k)^2$. The $\mathcal{O}(\epsilon^0)$ piece again vanishes on angular averaging, leaving
   \begin{align}
       P_{\Pi_E F}^{\rm 3h, (1)}(k_\perp) &= \left[\left(\frac{A_1}{3} + \frac{A_3}{5}\right) \frac{\bU{\rm g}(k)}{k} - \left(\frac{A_0}{3} + \frac{A_2}{5}\right) \frac{\partial \bU{\rm g}(k)}{\partial k}\right] \nonumber \\[0.2em]
       &\times \frac{\bU{F}(k) \, P_{\rm lin}(k)}{2\pi^2} \int_0^{k_L} \mathrm{d}q \, q^2\, \bU{\rm g}(q)\, P_{\rm lin}(q). \label{eq:P_3h_1}
   \end{align}
   The residual three-halo contamination thus receives contributions from two distinct sources: the scale-dependence of the short-wavelength biased profile $\partial_k \bU{\rm g}$, and the mode coupling encoded in the coefficients $A_1$ and $A_3$. The latter combines two effects entering at the same order in $\epsilon$: the expansion of the $F_2$ kernels, and that of the linear power spectrum about $k$, which contributes the slope $n_k$ and the curvature $\alpha_k$. Both are suppressed by a power of $\epsilon \equiv q/k$ and, since the filter enforces $q < k_L \ll k$, the residual is parametrically small. Appendix~\ref{app:beyond_squeezed} verifies that Equation~\eqref{eq:A_i} reproduces the known matter-only squeezed-limit response of \cite{2212.11940} at $\mathcal{O}(\epsilon^0)$.
   
   In summary, all three halo terms behave identically at leading order with the exact squeezed-limit contribution vanishing due to the angular kernel being odd in $\mu$ and the first non-zero correction appearing only at $\mathcal{O}(q/k)$. The amplitude of the residual is governed by the scale-dependence of the halo model kernels and by the logarithmic slope and curvature of the linear matter power spectrum. This is the analytical counterpart of the suppression reported in \cite{Hotinli2026}, and Equations~\eqref{eq:P_1h_1}, \eqref{eq:P_2h_1}, and \eqref{eq:P_3h_1} provide a template that can be included in a model when the residual becomes relevant at higher significance.

\section{Full-sky generalisation} \label{sec:full_sky}
  The derivations presented in the previous sections were performed within the flat-sky limit. While this provides considerable analytical insight, and is likely valid in the small-scale limit where the ML signal may be reliably measured, a full curved-sky derivation may reveal interesting residual contributions that could become relevant in future high-sensitivity experiments. We thus extend the previous formalism by expanding the projected fields in spherical harmonics. The moving lens temperature and foreground fields are scalar quantities and are naturally described by ordinary spherical harmonics, whereas the projected galaxy momentum is a spin-1 field and is therefore expanded in spin-weighted spherical harmonics. We derive the corresponding harmonic coefficients before constructing the full-sky angular power spectra. Here, we quote the results, with the full algebra given in Appendix~\ref{app:spin-1}.

 \subsection{Scalar fields}
  Consider a projected scalar field,
  \begin{equation}
      \lambda(\nv) = \int\mathrm{d}\chi\, W_\lambda(\chi)\, \Lambda(\chi\nv).
  \end{equation}
  Expanding this field in Fourier space, applying the plane-wave expansion, and using the orthogonality of the spherical harmonics gives
  \begin{equation}
      \lambda_{\ell m} = \int\frac{\mathrm{d}^3 k}{(2\pi)^3}\,\lambda_{\ell m}({\bf k}),\hspace{12pt}\lambda_{\ell m}({\bf k})\equiv4\pi i^\ell \int\mathrm{d}\chi\,  W_{\lambda}(\chi)\, \Lambda(\mathbf{k})\, j_\ell(k\chi)\, Y_{\ell m}^*(\hat{\mathbf{k}}).
  \end{equation}
  Throughout this section, when working with Fourier-space quantities such as $\lambda_{\ell m}({\bf k})$, we adopt the frame in which $\hat{\mathbf{k}}=\hat{\mathbf{z}}$ without loss of generality, since the final power spectra are rotationally invariant. As shown in \cite{1301.3607,2509.18732} in the context of the kSZ effect, this choice significantly simplifies the curved-sky treatment of fields involving radial or transverse projections of vector quantities. In this frame, we may write
  \begin{equation}
      Y_{\ell m}^*(\hat{\mathbf{z}}) = \sqrt{\frac{2\ell+1}{4\pi}}\,\delta_{m,0},
  \end{equation}
  so that
  \begin{equation} \label{eq:lambda_lm}
      \lambda_{\ell m}({\bf k}) = \delta_{m,0} i^\ell \sqrt{4\pi(2\ell+1)} \int\mathrm{d}\chi\, W_{\lambda}(\chi)\, \Lambda(\mathbf{k})\, j_\ell(k\chi).
  \end{equation}
  The same result holds for the foreground field $\delta_{f,\ell m}$, replacing $W_\lambda \to W_f$. Hence, any scalar field constructed from the radial projection of a three-dimensional scalar, regardless of its physical origin, has support only at $m=0$ in the frame aligned with $\hat{\mathbf{k}}$.

  As discussed in Section \ref{ssec:ml.definition}, strictly speaking, the ML effect depends only on the transverse components of the peculiar velocity and the gradient of the Newtonian potential, and is therefore not the projection of a three-dimensional scalar. Although the radial contribution is not parametrically suppressed relative to the transverse one and would contribute to an ISW- or Rees--Sciama-type calculation without suppression, its correlation with the transverse galaxy momentum is suppressed. As shown in Appendix~\ref{app:radial}, the angular structure of the estimator removes the leading contribution, leaving a residual of $\mathcal{O}(q^2/k^2)$. Thus, treating the ML source as a three-dimensional scalar is justified in precisely the regime in which the estimator is applied.

 \subsection{The spin-1 momentum field} \label{ssec:full_sky.spin-1}
  As before, we build the projected galaxy momentum field by decomposing the three-dimensional momentum field into longitudinal and transverse modes $\mathbf{q}({\bf k})=q^{\rm L}({\bf k})\hat{\bf k}+\mathbf{q}^{\rm T}({\bf k})$, as in Equation~\eqref{eq:perp_par}, and applying the projector $\mathcal{P}(\nv) = \mathbb{I}-\nv\nv^\dagger$ to each piece. We can then define a complex spin-1 field by contracting the resulting vector quantity with the polarisation basis,
  \begin{equation}
      \pi(\nv) = \boldsymbol{\pi}(\nv)\cdot\hat{\bf e}_\perp, \quad\quad \hat{\bf e}_\perp\equiv\hat{\mathbf{e}}_\theta + i\hat{\mathbf{e}}_\varphi = -\eth\nv.
  \end{equation}
  Here $\hat{\bf e}_\theta\propto\partial_\theta\hat{\bf n}$ and $\hat{\bf e}_\varphi\propto\partial_\varphi\hat{\bf n}$ are the unit vectors in the directions of the spherical coordinates $\theta$ and $\varphi$, and $\eth$ is the spin-raising differential operator (see \cite{1967JMP.....8.2155G} for a full description of the $\eth$ formalism and spin-weighted spherical harmonics).

  Writing the polarisation vector as a spin-raising derivative of $\nv$ allows the angular integrals to be performed by parts and reduced to Gaunt integrals. Since $\nv\cdot\hat{\bf e}_\perp=0$, the term $\mathcal{P}\cdot\mathbf{q}^{\rm T}$ proportional to $\nv$ drops out. The $E$- and $B$-modes follow from
  \begin{equation} \label{eq:EB_full_sky_def}
      \pi_{\ell m} = -\int\mathrm{d}\hat{\mathbf{n}}\, _1Y_{\ell m}^*(\hat{\mathbf{n}})\, \pi(\hat{\mathbf{n}}), \quad\quad \pi_{\ell m}^E = \frac{1}{2}(\pi_{\ell m} + \bar{\pi}_{\ell m}), \quad\quad i\pi_{\ell m}^B = \frac{1}{2}(\pi_{\ell m} - \bar{\pi}_{\ell m}),
  \end{equation}
  where the complex-conjugate field $\bar{\pi}$ is
  \begin{equation}
    \bar{\pi}\equiv\boldsymbol{\pi}(\nv)\cdot\hat{\bf e}^*_\perp,\hspace{12pt}\bar{\pi}_{\ell m}=-\int d\nv\,_{-1}Y_{\ell m}^*(\nv)\,\bar{\pi}(\nv),
  \end{equation}
  with $\hat{\bf e}^*_\perp\equiv\hat{\bf e}_\theta-i\hat{\bf e}_\varphi=-\bar{\eth}\nv$, and $_sY_{\ell m}(\nv)$ are the spin-$s$ spherical harmonic functions.
  
  We first consider the longitudinal component. Carrying out the harmonic decomposition (Appendix~\ref{app:spin-1}), one finds $\bar{\pi}_{\ell m}^{\rm L} = \pi_{\ell m}^{\rm L}$, and hence
  \begin{equation}
      \pi_{\ell m}^{{\rm L}, B} = 0, \quad\quad \pi_{\ell m}^{{\rm L},E} = \pi_{\ell m}^{\rm L}.
  \end{equation}
  The longitudinal momentum is therefore a pure gradient field on the curved sky. Evaluating the relevant Wigner-$3j$ coefficients arising from the angular integrals over three spherical harmonic functions, and applying the recursion relation of the spherical Bessel functions, we obtain the result
  \begin{equation} \label{eq:pi_lm_par_E}
      \pi_{\ell m}^{{\rm L},E} = \delta_{m,0}\,i^{\ell-1} \sqrt{4\pi(2\ell+1)} \int\mathrm{d}\chi\, W_{\pi}(\chi) \int\frac{\mathrm{d}^3 k}{(2\pi)^3}\, q^{\rm L}(\mathbf{k},\chi)\, \frac{\sqrt{\ell(\ell+1)}}{k\chi}j_{\ell}(k\chi).
  \end{equation}
  Hence, the longitudinal $E$-mode has exactly the same azimuthal structure as a scalar field, leading to a non-zero correlation between the two.

  We next consider the transverse component. Orientating $\mathbf{q}^{\rm T} = q_\perp\hat{\mathbf{y}}$ in the same frame, the factor $\nv\cdot\mathbf{q}^{\rm T} \propto \sin\theta\sin\varphi$ is a pure dipole. Combined with the Wigner selection rules, this enforces $m=\pm1$, and gives
  \begin{equation} \label{eq:pi_lm_perp_E}
      \pi_{\ell m}^{{\rm T},E} = \delta_{m,\pm1}\,i^{\ell+1} \sqrt{4\pi(2\ell+1)} \int\mathrm{d}\chi\, W_{\pi}(\chi) \int\frac{\mathrm{d}^3 k}{(2\pi)^3}\, \mathbf{q}^{\rm T}(\mathbf{k},\chi)\, \frac{1}{2}\left(j_\ell'(k\chi) + \frac{j_{\ell}(k\chi)}{k\chi}\right),
  \end{equation}
  \begin{equation} \label{eq:pi_lm_perp_B}
      \pi_{\ell m}^{{\rm T},B} = \delta_{m,\pm1}\,i^{\ell+1} \sqrt{4\pi(2\ell+1)} \int\mathrm{d}\chi\, W_{\pi}(\chi) \int\frac{\mathrm{d}^3 k}{(2\pi)^3}\, \mathbf{q}^{\rm T}(\mathbf{k},\chi)\, \frac{j_\ell(k\chi)}{2}.
  \end{equation}
  The transverse component therefore generates both $E$- and $B$-modes, but both are supported only at $m=\pm1$. Hence, the cross-correlation between the transverse component and a scalar field vanishes exactly on all scales. As a result, only the longitudinal component, being a pure $E$-mode with $m=0$, survives in the cross-correlation. That the $B$-mode correlates with neither the signal nor the foregrounds is also what parity requires, and the calculation makes this explicit rather than assumed.

 \subsection{Angular power spectra}
  Combining Equations~\eqref{eq:lambda_lm} and \eqref{eq:pi_lm_par_E}, the diagonal form
  \begin{equation}
      \left\langle \pi_{\ell m}^{{\rm L},E} \lambda_{\ell'm'}^* \right\rangle = \delta_{\ell\ell'}\, \delta_{mm'}\, C_\ell^{\pi_E \lambda}
  \end{equation}
  enforces that $\ell'=\ell$, giving
  \begin{equation} \label{eq:Cl_pi_lambda_full}
      C_\ell^{\pi_E \lambda} = \frac{2}{\pi} \int\mathrm{d}\chi\, W_{\pi}(\chi) \int\mathrm{d}\chi'\, W_{\lambda}(\chi') \int\mathrm{d}k\,k^2\, P_{q^{\rm L} \Lambda}(k;\chi,\chi')\, \frac{\sqrt{\ell(\ell+1)}}{k\chi}\, j_{\ell}(k\chi)j_{\ell}(k\chi'),
  \end{equation}
  where
  \begin{equation}
      \left\langle q^{\rm L}(\mathbf{k},\chi) \, \Lambda^*(\mathbf{k}',\chi') \right\rangle = (2\pi)^3 \, \delta^{\mathcal{D}}(\mathbf{k}-\mathbf{k}') \, P_{q^{\rm L} \Lambda}(k;\chi,\chi').
  \end{equation}
  The longitudinal momentum in Fourier space is
  \begin{equation}
      q^{\rm L}(\mathbf{k}) = iaHf\int\frac{\mathrm{d}^3 p}{(2\pi)^3}\, \Theta(|\mathbf{k}-\mathbf{p}|)\, \frac{\hat{\mathbf{k}}\cdot(\mathbf{k}-\mathbf{p})}{|\mathbf{k}-\mathbf{p}|^2}\, \delta_{\rm g}(\mathbf{p})\, \delta_{\rm g}(\mathbf{k}-\mathbf{p}).
  \end{equation}
  
  The disconnected cross-spectrum between $q^{\rm L}({\bf k})$ and $\Lambda({\bf k})$ is then given by
  \begin{equation}
      P_{q^{\rm L} \Lambda}^{\rm (dc)}(k) = aHf \int\frac{\mathrm{d}^3 q}{(2\pi)^3}\, \Theta(q)\, \frac{\mathbf{q}\cdot\mathbf{k}}{q^2k}\; \frac{\mathbf{q}\cdot(\mathbf{k}-\mathbf{q})}{q^2|\mathbf{k}-\mathbf{q}|^2}\, \mathcal{S}(\mathbf{q},\mathbf{k}-\mathbf{q}),
  \end{equation}
  where
  \begin{equation} \label{eq:calS}
      \mathcal{S}(\mathbf{q},\mathbf{k}-\mathbf{q}) \equiv P_{\mathrm{g}v}(q)\,P_{\mathrm{gm}}(|\mathbf{k}-\mathbf{q}|) + P_{\mathrm{g}v}(|\mathbf{k}-\mathbf{q}|)\,P_{\mathrm{gm}}(q)
  \end{equation}
  collects the two Wick contractions and retains the asymmetry between them established in Section~\ref{sssec:ml.power_spectrum_disconnected}. In the squeezed limit, the two factors of $\mathbf{q}\cdot\hat{\mathbf{k}}$ combine into $\mu^2$, with angular average $\langle\mu^2\rangle=1/3$, giving
  \begin{equation} \label{eq:P_qLambda_sq}
      P_{q^{\rm L} \Lambda}^{\rm (dc)}(k) \simeq \frac{aHf}{3k} \left[P_{\mathrm{gm}}(k)\,\mathcal{A}_v(k_L) + P_{\mathrm{g}v}(k)\,\mathcal{A}_{\rm m}(k_L)\right],
  \end{equation}
  with $\mathcal{A}_X$ defined in Equation~\eqref{eq:A_def}. This is the curved-sky counterpart of Equation~\eqref{eq:P_dc_total}, and makes the effect of the filter on the signal explicit. The $k_L$ cut enters only through the filtered variances $\mathcal{A}_X$. These are dominated by modes near the peak of the linear power spectrum but converge only slowly above it, so the signal retains a weak residual dependence on $k_L$, which we quantify in Section~\ref{ssec:results.signal_vs_fgs}.

  Similarly, repeating the calculation for the foreground field gives the same structure with $W_\lambda \to W_f$ and 
  \begin{equation}
      P_{q^{\rm L} F}(k;\chi,\chi') = aHf\int\frac{\mathrm{d}^3 q}{(2\pi)^3}\, \Theta(q)\, \frac{\hat{\mathbf{k}}\cdot\mathbf{q}}{q^2}\, B_{\mathrm{gg}F}(\mathbf{k}-\mathbf{q},\mathbf{q},-\mathbf{k};\chi,\chi').
  \end{equation}
  In the squeezed limit, the response form $B_{\mathrm{gg}F} \simeq P_{\rm gm}(q) P_{F\mathrm{g}}(k)\,{\cal R}_{F\mathrm{g}}(k)$ again gives an angular integrand odd in $\mu$, and hence $P_{q^{\rm L} F} \to 0$ as $q/k \to 0$. The cancellation is therefore not an artefact of the flat-sky or Limber approximations, but rather a consequence of the angular structure of the squeezed bispectrum that survives the full curved-sky treatment.
  
 \subsection{Recovery of the flat-sky limit}
  In the Limber limit, $j_\ell(k\chi) \to \sqrt{\pi/(2\ell+1)}\, \delta^{\mathcal{D}}(\ell+1/2-k\chi)$, and Equation~\eqref{eq:Cl_pi_lambda_full} collapses into Equation~\eqref{eq:Cl_pi_lambda_flat} up to the factor $\sqrt{\ell(\ell+1)}/(k\chi)\simeq\sqrt{\ell(\ell+1)}/(\ell+1/2) \to 1$. This confirms that the full-sky expression reproduces the flat-sky calculation on small angular scales.

\section{Numerical results} \label{sec:results}
  We now evaluate some of the expressions derived in the previous sections using the halo model to describe the galaxy, matter, velocity, and foreground fields. We focus particularly on the behaviour of the ML and foreground terms as a function of the measurement scale $k_\perp$ and the filtering scale $k_L$ used in velocity reconstruction.
  
 \subsection{Implementation} \label{ssec:results.implementation}
   We implement the halo model expressions using the Core Cosmology Library (CCL) \cite{Chisari2019core}, adopting a halo mass definition with overdensity threshold $\Delta=200$ with respect to the critical density, the concentration-mass relation of \cite{Duffy2008dark}, the halo mass function of \cite{Tinker2008toward}, and the halo bias of \cite{Tinker2010large}. We assume a fiducial cosmology consistent with \textit{Planck} \cite{Planck2018results},
   \begin{equation}
     \{\Omega_{\rm c},\, \Omega_{\rm b},\, h,\, n_{\rm s},\, \sigma_8\} = \{0.2607,\, 0.04897,\, 0.6766,\, 0.9665,\, 0.8102\}.
   \end{equation}
   The matter overdensity is modelled with an NFW profile \cite{Navarro1997universal}. The galaxy overdensity uses the halo occupation distribution (HOD) implemented in CCL with
   \begin{equation}
      \{\log_{10} M_{\rm min},\, \log_{10} M_0,\, \log_{10} M_1,\, \alpha\} = \{12.89,\, 12.92,\, 13.95,\, 1.1\},
   \end{equation}
   compatible with a luminous red galaxy sample \cite{2309.06443}. All spectra are evaluated at $z=0.55$, matching the redshift at which the HOD parameters were measured. The foreground field is modelled as the tSZ effect, using the hydrostatic equilibrium pressure profile of \cite{Ferreira2023xray, LaPosta2025insights} with
   \begin{equation}
     \{\log_{10} M_{\rm c},\, \beta,\, \eta_{\rm b},\, A_*\} = \{14.0,\, 0.6,\, 0.5,\, 0.03\}.
   \end{equation}
   The electron pressure profile is converted to a temperature fluctuation through
   \begin{equation}
       \left.\frac{\delta T}{T}\right|_{\rm tSZ} = g(\nu)\,y, \quad\quad y = \frac{\sigma_{\rm T}}{m_{\rm e} c^2} \int\frac{\mathrm{d}\chi}{1+z}\, P_{\rm e},
   \end{equation}
   where $g(x) = x\coth(x/2)-4$ and $x=h\nu/(k_{\rm B} T_{\rm CMB})$. We quote results at $\nu = 150\,\mathrm{GHz}$, where $g(\nu) = -0.95$. The moving lens curves carry the corresponding kernel $W_\lambda = -3 H_0^2 \Omega_{\rm m} Hf$ from Equation~\eqref{eq:W_lambda}. Both spectra are therefore expressed as temperature fluctuations per unit comoving distance, in units of $\mathrm{Mpc}^3$.
   
   The logarithmic slope $n_k$, curvature $\alpha_k$, and profile derivative $\partial_k \bU{\rm g}$ are computed by centred finite differences with logarithmic step $\Delta\ln k = 10^{-3}$. Long-mode integrals are evaluated by Simpson integration on logarithmically spaced $q$-grids. We consider measurement scales $10^{-3} \leq k_\perp \leq 10 \, \mathrm{Mpc}^{-1}$, with the sampling refined over $6 \times 10^{-3} \leq k_\perp \leq 0.5\, \mathrm{Mpc}^{-1}$ in order to resolve the baryon acoustic features that enter through $n_k$ and $\alpha_k$. Finally, we consider velocity reconstruction scale cuts
  \begin{equation}
      k_L = \{0.10,\, 0.20,\, 0.50\}\, \mathrm{Mpc}^{-1}.
  \end{equation}

 \subsection{Structure of the moving-lens signal} \label{ssec:results.ml_signal}
  \begin{figure}[ht]
      \centering
      \includegraphics[width=0.8\linewidth]{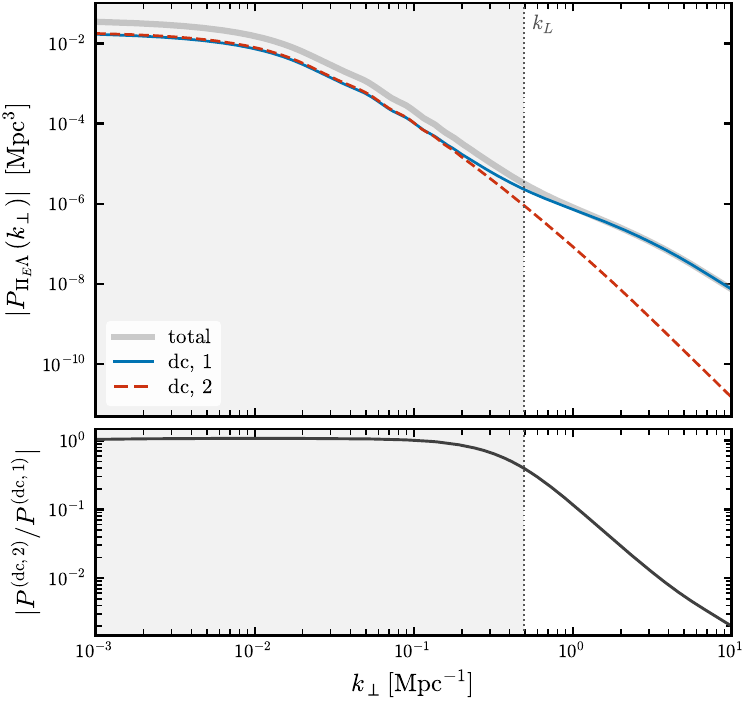}
      \caption{The two disconnected contributions to the moving-lens signal at fixed reconstruction cut $k_L = 0.5\,\mathrm{Mpc}^{-1}$ (vertical dotted line). The shaded region $k_\perp < k_L$ lies outside the squeezed regime. \emph{Top:} $P^{\rm (dc,1)}_{\Pi_E \Lambda}$ (solid blue) and $P^{\rm (dc,2)}_{\Pi_E \Lambda}$ (dashed red), together with their sum (grey). \emph{Bottom:} the ratio of the two.}
      \label{fig:dc1_dc2}
  \end{figure}
  
  Before comparing the moving-lens signal with the foregrounds, we first examine the structure of the signal itself. As discussed in Section~\ref{sssec:ml.power_spectrum_disconnected}, the disconnected part of the four-point function contains two non-vanishing Wick contractions which would be identical for a symmetric estimator. Here, however, the two contractions are distinguished by the construction of the velocity template: the reconstruction contains no one-halo power, so $P_{\mathrm{g}v}$ retains only the two-halo contribution, whereas $P_{\mathrm{gm}}$ retains the full non-linear power.
  
  Figure~\ref{fig:dc1_dc2} shows the two contractions separately, together with their sum. On large scales, the two are indistinguishable since non-linear effects are negligible there, and the total is simply twice either contraction. The second contraction is suppressed on smaller scales, since its short-wavelength factor is $P_{\mathrm{g}v}$, which by construction has no one-halo contribution. The ratio of the two departs from unity near $k_\perp \sim 0.1\,\mathrm{Mpc}^{-1}$ and has fallen to roughly $3 \times 10^{-3}$ by $k_\perp = 10\,\mathrm{Mpc}^{-1}$, so that the small-scale signal is carried almost entirely by the first contraction.

 \subsection{Foreground suppression versus the moving-lens signal} \label{ssec:results.signal_vs_fgs}
\begin{figure}[ht]
      \centering
      \includegraphics[width=\linewidth]{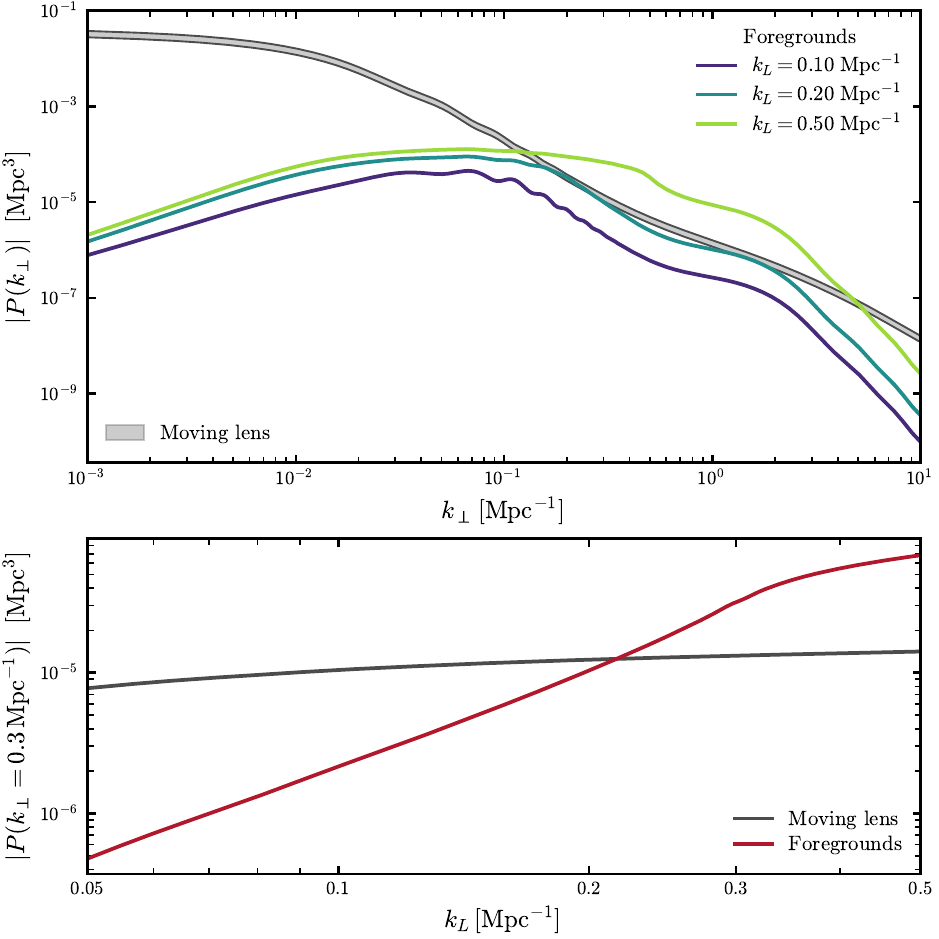}
      \caption{Comparison of total foreground contamination and moving-lens signal. \emph{Top:} Dependence on transverse wavenumber. The moving-lens signal is shown as a grey band spanning the full range of reconstruction cuts, while the total foreground power spectrum is shown as a separate coloured curve for each cut. Both quantities are shown as temperature fluctuations per unit comoving distance, in $\mathrm{Mpc}^3$, at $z=0.55$ and $\nu = 150\,\mathrm{GHz}$. \emph{Bottom:} The same quantities evaluated at fixed $k_\perp = 0.3\,\mathrm{Mpc}^{-1}$ as a function of the reconstruction cut $k_L$.}
      \label{fig:signal_vs_fgs}
  \end{figure}

  Figure~\ref{fig:signal_vs_fgs} compares the total foreground contamination with the moving-lens signal in two complementary ways. The upper panel shows both as a function of the transverse wavenumber $k_\perp$: the moving-lens signal is shown as a grey band spanning the full range of reconstruction cuts $k_L\in(0.1,0.5)\,{\rm Mpc}^{-1}$, while the total foreground is shown as a separate curve for each cut scale. The lower panel fixes the measurement scale at $k_\perp = 0.3\,\mathrm{Mpc}^{-1}$ and shows the dependence of both quantities on the reconstruction cut $k_L$, making their relative scaling with $k_L$ immediately apparent.

  The central result is the contrast in how the signal and the foregrounds respond to $k_L$. The moving lens contribution is nearly insensitive to the reconstruction cut: the grey band in the upper panel of Figure~\ref{fig:signal_vs_fgs} is narrow and its shape is unchanged. This is a consequence of $k_L$ entering only through an overall multiplicative amplitude in the squeezed limit, rather than through the spectral shape, as shown in Section~\ref{ssec:ml.power_spectrum}. The foreground curves, by contrast, separate systematically and in the expected direction, with tighter cuts reducing the contamination at all transverse scales. The lower panel quantifies this behaviour. Over the range of $k_L$ considered, the moving-lens signal changes by a factor of $1.75$ while the total foreground changes by a factor of $\sim120$. Restricting an analysis to smaller $k_L$ therefore removes foreground power at a substantially faster rate than it removes the signal. Moreover, the two curves cross at $k_L \simeq 0.21\,\mathrm{Mpc}^{-1}$: for tighter cuts the residual tSZ contamination at $k_\perp = 0.3\,\mathrm{Mpc}^{-1}$ lies below the moving-lens signal.

  Two caveats apply to the absolute normalisation. First, the foreground amplitude is proportional to $g(\nu)$, which vanishes near $217\,\mathrm{GHz}$ and approaches $|g|\simeq2$ in the Rayleigh--Jeans limit. Hence, the curves shown correspond to a single observing frequency. Second, no component separation has been applied, so the curves represent the full tSZ field rather than a realistic residual foreground. Both choices make the comparison conservative, while neither affects the relative scaling with $k_L$, which is the focus of the figure.

\begin{figure}[ht]
      \centering
      \includegraphics[width=0.7\linewidth]{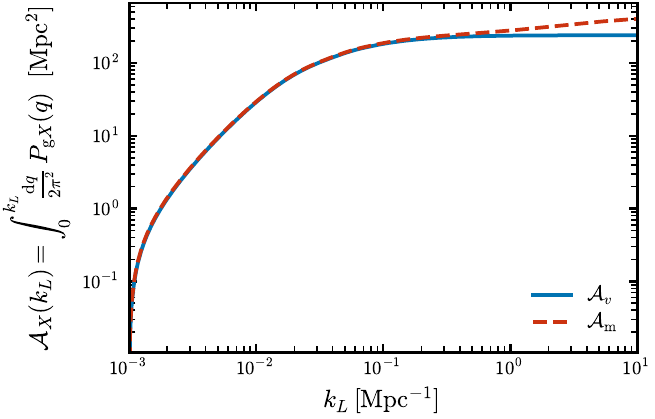}
      \caption{The filtered variances $\mathcal{A}_X(k_L)$ of Equation~\eqref{eq:A_def}, through which the reconstruction cut enters the disconnected moving-lens signal: $\mathcal{A}_v$ (solid blue), which multiplies the dominant Wick contraction, and $\mathcal{A}_{\rm m}$ (dashed red), which multiplies the subdominant contraction.}
      \label{fig:A_kL}
  \end{figure}

  The two filtered variances defined in Equation~\eqref{eq:A_def} are shown explicitly in Figure~\ref{fig:A_kL}. On large scales, where non-linear effects are negligible and $P_{\mathrm{g}v} \simeq P_{\mathrm{gm}}$, the two are indistinguishable, and only begin to separate above $k_L \simeq 0.2\,\mathrm{Mpc}^{-1}$. The amplitude $\mathcal{A}_v$, which multiplies the dominant contraction, saturates by $k_L \simeq 0.4\,\mathrm{Mpc}^{-1}$, as the velocity template contains no one-halo power so $P_{\mathrm{g}v}$ falls steeply on small scales and the integral converges. The amplitude $\mathcal{A}_{\rm m}$, which multiplies the subdominant contraction, continues to grow, since $P_{\mathrm{gm}}$ retains the full non-linear power. This saturation is the origin of the weak residual dependence of the signal on $k_L$ seen in the lower panel of Figure~\ref{fig:signal_vs_fgs}.

\begin{figure}[ht]
      \centering
      \includegraphics[width=\linewidth]{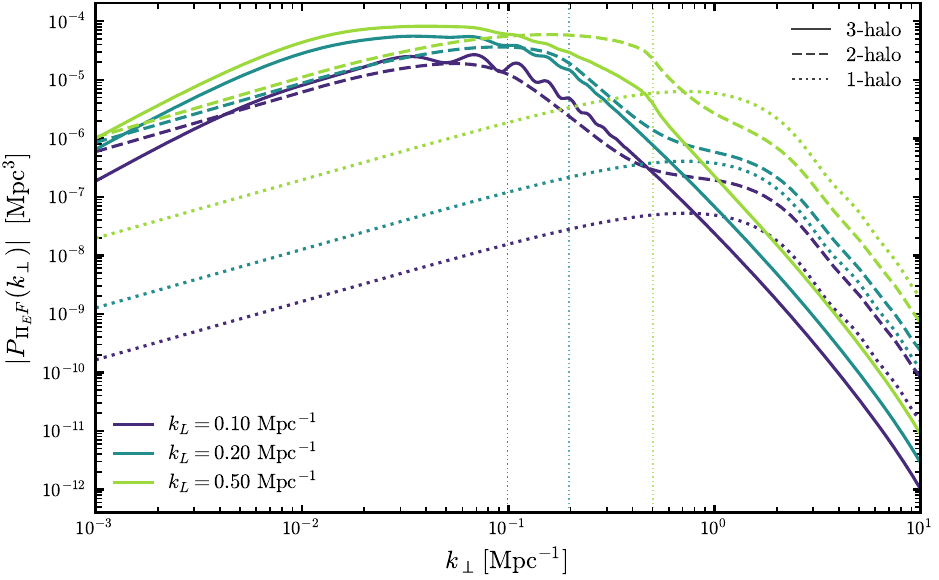}
      \caption{Halo model decomposition of the foreground residual: three-halo (solid), two-halo (dashed), and one-halo (dotted) contributions for each reconstruction cut (colours). Vertical dotted lines mark the corresponding values of $k_L$.}
      \label{fig:halo_decomposition}
  \end{figure}

  Figure~\ref{fig:halo_decomposition} breaks the total foreground of Figure~\ref{fig:signal_vs_fgs} into its one-, two-, and three-halo contributions. The individual terms respond in qualitatively different ways. The three-halo term dominates the foreground budget on large scales, peaking near $k_\perp \sim 0.03$--$0.08\,\mathrm{Mpc}^{-1}$, before falling steeply once $k_\perp$ exceeds the cut. The two-halo term takes over at intermediate scales, and the one-halo term rises steadily throughout, so that by $k_\perp = 10\,\mathrm{Mpc}^{-1}$ the three-halo term is the smallest of the three. For the widest cut, $k_L = 0.5\,\mathrm{Mpc}^{-1}$, the one-halo term exceeds the two-halo term on the smallest scales whereas, for the tighter cuts, the two-halo term remains the largest contribution.

  The three terms are also suppressed by the reconstruction scale cut in different ways. The one-halo suppression is scale independent: the cut reduces this term by the same factor at every $k_\perp$, and this factor is very close to $(k_L/k_L^{\rm max})^3$. This is what one expects from Equation~\eqref{eq:P_1h_1}: when the soft mode lies well outside the halo scale, the profile $u_{\rm g}(q|M)$ is essentially constant over the range of integration and the residual $k_\perp$ dependence factorises out entirely. It is the same flatness of the profile that we invoke in Section~\ref{ssec:results.squeezed_validity} to explain why the one-halo beyond-squeezed expression is so accurate. The two- and three-halo suppressions, by contrast, are strongly scale dependent, being weakest on the largest and smallest scales and strongest for transverse modes comparable to the widest reconstruction cut.
  
 \subsection{Validity of the beyond-squeezed expansion} \label{ssec:results.squeezed_validity}
  \begin{figure}[ht]
      \centering
      \includegraphics[width=\linewidth]{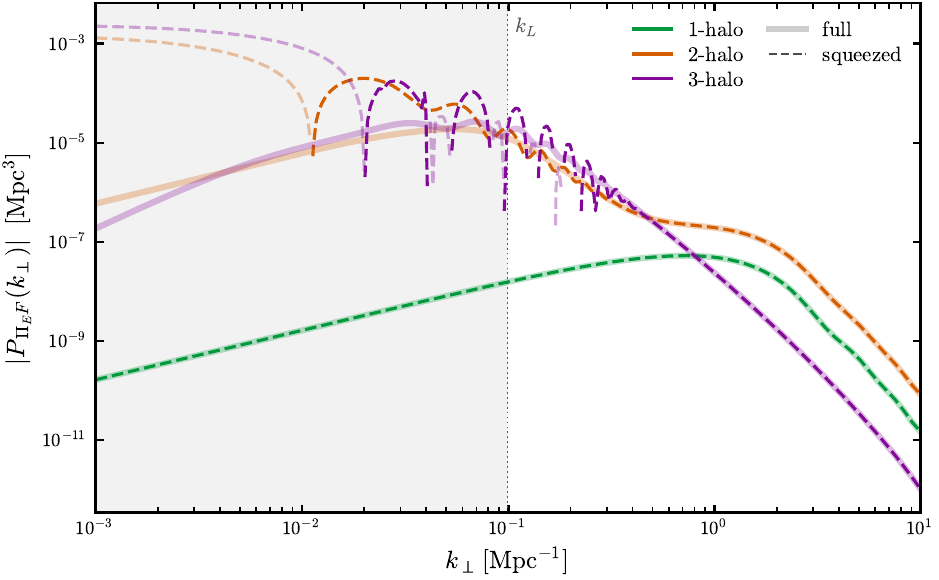}
      \caption{Validity of the leading beyond-squeezed expansion at fixed reconstruction cut $k_L = 0.1\,\mathrm{Mpc}^{-1}$ (vertical dotted line). The full halo-model foreground power spectrum $P_{\Pi_E F}(k_\perp)$ (pale band) is compared with the leading $\mathcal{O}(q/k)$ expressions of Section~\ref{ssec:foregrounds.beyond_squeezed} (dashed), for the one-halo (green), two-halo (orange), and three-halo (purple) contributions. Curves show $|P|$, with negative stretches drawn at reduced opacity, so the cusps in the three-halo expansion below $k_\perp \simeq 0.3\,\mathrm{Mpc}^{-1}$ are sign changes. The shaded region marks $k_\perp < k_L$.}
      \label{fig:squeezed}
  \end{figure}
  
  Figure~\ref{fig:squeezed} compares the full halo model calculation with the leading beyond-squeezed expressions of Section~\ref{ssec:foregrounds.beyond_squeezed} at fixed $k_L = 0.1\,\mathrm{Mpc}^{-1}$. For $k_\perp \gtrsim 0.3\,\mathrm{Mpc}^{-1}$, the three-halo expansion reproduces the full calculation closely, tracking it over more than a decade and a half in $k_\perp$. This is the quantitative statement behind the analytical argument of Section~\ref{ssec:foregrounds.squeezed_limit}: the exact squeezed contribution cancels identically, and what survives is the first correction in $q/k$.
  
  The beyond-squeezed one-halo term, on the other hand, is indistinguishable from the full calculation over the entire range of $k_\perp$, including well below $k_L$. This is because the one-halo bispectrum depends only on the magnitudes of the wavevectors and, on the scales of interest here (i.e. well below the virial wavenumber, where the halo profiles are essentially flat), the profile $u(|\mathbf{k}-\mathbf{q}|)$ can be linearised about its zero-argument value. The residual angular integral then depends only on the ratio $q/k$ and coincides with the value it takes in the squeezed limit. The full and beyond-squeezed expressions are therefore close to equivalent, which also accounts for the exactly scale-independent suppression of the one-halo term observed in Figure~\ref{fig:halo_decomposition}.

  The two-halo term, shown in orange in Figure~\ref{fig:squeezed}, provides a useful control on this interpretation. Its beyond-squeezed expression, given by Equation~\eqref{eq:P_2h_1}, involves no derivative of the linear power spectrum beyond its logarithmic slope $n_k$; unlike the three-halo term, it carries no dependence on the curvature $\alpha_k$. It converges to the full calculation from $k_\perp \simeq 2k_L$, whereas the three-halo term continues to oscillate until $k_\perp \simeq 3k_L$, and the one-halo term (which involves no linear power spectrum at all) agrees at every scale shown. The ordering of these three convergence scales follows from the number of derivatives of $P_{\rm lin}$ that each term requires.

\section{Conclusions} \label{sec:conclusions}
  We have examined the mathematical structure of the harmonic-space moving-lens estimator of \cite{Hotinli2026}, consisting of the cross-correlation between a CMB map and the $E$-mode of the projected galaxy momentum field $\boldsymbol{\pi}_{\rm g}$ \cite{2605.15947}, with the galaxy velocities reconstructed after nulling all small-scale modes $k\gtrsim k_L$. We have presented expressions for the harmonic components of both fields, including both $E$- and $B$-mode components for $\boldsymbol{\pi}_{\rm g}$, as well as their cross-correlation, and the correlation between $\boldsymbol{\pi}_{\rm g}$ and residual extragalactic foreground contamination. These results have been derived exactly in the curved sky, as well as in the flat-sky and Limber approximations.

  A focus of this work has been to provide an analytical explanation for the effectiveness of the $k_L$ cut at suppressing the contribution from extragalactic foregrounds, as demonstrated empirically in \cite{Hotinli2026}. This suppression follows from symmetry together with the nature of gravitational collapse. On small scales, the galaxy momentum field depends linearly on the direction cosine $\mu$ of the reconstructed velocity mode, a structure mirrored by the ML contribution itself. Their cross-correlation is therefore a trispectrum integrated over squeezed configurations and weighted by $\mu^2$, which depends on $k_L$ only through the large-scale velocity variance. Foreground contamination carries no such $\mu$ dependence, and its cross-correlation with $\boldsymbol{\pi}_{\rm g}$ is a bispectrum, integrated over the same configurations but weighted by a single power of $\mu$. Since the squeezed large-scale-structure bispectrum is independent of the soft-mode direction, this integral vanishes. The cancellation holds for any foreground that can be written as an angular projection of a statistically isotropic three-dimensional tracer, independent of its SED or spatial profile, and does not rely on the flat-sky or Limber approximations. 
  
  The filter therefore affects the ML signal only through an overall normalisation set by the filtered velocity variance. Since this variance converges soon after the peak of the linear matter power spectrum, the signal depends only weakly on the filtering scale, whereas foreground contamination falls by a factor of $\sim120$ over the range of cuts considered ($k_L\in(0.05,\,0.5)\,{\rm Mpc}^{-1}$).

  The same behaviour may also be explained physically. The ML cross-correlation is carried by the large-scale bulk velocity field, which dominates both the ML signal and the galaxy momentum $E$-mode. By contrast, the correlation with foreground structures arises from the correspondence between the local velocity field and the gradient of the gravitational potential in which these structures reside. The $k_L$ filter erases this local velocity component (and, with it, the foreground correlation), leaving the large-scale flow that galaxies and foregrounds share and that sources the ML effect.
 
  We then computed the leading foreground residual beyond the exact squeezed limit within the halo model. The one-, two-, and three-halo contributions all first appear at $\mathcal{O}(q/k)$, with amplitudes set by the scale dependence of the short-wavelength halo profiles and the logarithmic slope and curvature of the linear matter power spectrum. Since the squeezed-limit contribution vanishes identically, these terms constitute the complete leading residual rather than corrections to a non-zero result, and Equations~\eqref{eq:P_1h_1}, \eqref{eq:P_2h_1}, and \eqref{eq:P_3h_1} therefore provide a physically motivated template for it. A full halo-model evaluation confirms this picture: the beyond-squeezed expressions reproduce the exact calculation for $k_\perp \gtrsim 3k_L$, the regime carrying most of the estimator signal. The residual suppression is strongly scale-dependent for the two- and three-halo terms, whereas for the one-halo term it is independent of $k_\perp$, being set by the flatness of the galaxy profile across the filtered range. For current measurements, the residual lies comfortably below the statistical uncertainty, so $k_L$-filtering may be regarded as removing foregrounds entirely. As the precision of ML measurements improves, however, this will cease to hold and future analyses will be able to marginalise over the template derived here rather than assuming foregrounds vanish.

  Extending the framework to the full sky, we find that in a basis aligned with the Fourier wavevector, the momentum component longitudinal to $\mathbf{k}$ is a pure $E$-mode with $m=0$, while the perpendicular component carries both $E$- and $B$-modes but only at $m=\pm1$. Since scalar fields on the sky have only $m=0$, the vector mode decouples exactly from both the moving-lens signal and foreground contamination. The same treatment quantifies the error incurred by writing the ML effect as the full scalar product $\mathbf{v} \cdot \boldsymbol{\nabla}\phi$, rather than one involving only transverse components $\mathbf{v}_\perp \cdot \boldsymbol{\nabla}_\perp \phi$. As shown in Appendix~\ref{app:radial}, the radial contribution is removed in the squeezed limit by the same mechanism that removes the foreground bispectrum, and enters at $\mathcal{O}(q^2/k^2)$ rather than $\mathcal{O}(q/k)$. The ML signal may therefore be safely treated as a three-dimensional scalar product wherever such a simplification is analytically convenient.

  With the moving-lens signal only recently detected, several extensions to the analytical study presented here remain. For instance, the contribution of vortical velocity modes should be quantified. While this component should not correlate with  the momentum $B$-mode in a parity-conserving theory, the $E$-mode receives a connected contribution involving the vorticity trispectrum. Evaluating this contribution may require a helicity-based treatment such as that of \cite{Coulton2023}. More generally, the cancellation derived here is a statement about bispectrum shapes rather than about foregrounds specifically: any contribution whose squeezed limit is independent of the orientation of the soft mode is removed by the estimator, while one carrying a genuine dipolar dependence on $\mu$ would survive. Applying this criterion to primordial non-Gaussianity would establish the shapes, if any, to which the ML estimator is sensitive.
  
  Finally, we have made no attempt to account for non-linear contributions to the velocity field, assuming throughout that the galaxy--velocity power spectrum is sourced entirely by large-scale modes. This assumption breaks the symmetry between the two Gaussian contractions of Section~\ref{sssec:ml.power_spectrum_disconnected} and causes the filtered signal amplitude to saturate at small filtering scales. Characterising intra-halo velocities would be needed to quantify the connected trispectrum contribution to the signal. Whether that contribution is best regarded as part of the more general Rees--Sciama effect is partly a matter of convention, but its observational impact on ML searches remains to be established. This lies beyond the scope of the analytical framework developed here and would be addressed most naturally using hydrodynamical simulations, as has recently been done for the kSZ effect \cite{Ondaro-Mallea2026}. Such studies will become necessary as the sensitivity of CMB datasets increases and measurements of the ML signal become more precise.

\acknowledgments
 AW is supported by a STFC studentship, DA acknowledges support from the Beecroft Trust, and WRC is supported by an STFC Ernest Rutherford Fellowship. We made extensive use of computational resources at the University of Oxford Department of Physics, funded by the John Fell Oxford University Press Research Fund.

\appendix
 \section{Beyond-squeezed expansions} \label{app:beyond_squeezed}
  This appendix collects the squeezed-limit expansions of the halo-model bispectrum used in Section~\ref{ssec:foregrounds.beyond_squeezed}. Throughout we work in the configuration $(\mathbf{q}, \mathbf{k}-\mathbf{q},-\mathbf{k})$ and define $\epsilon \equiv q/k \ll 1$, $\mu \equiv \hat{\mathbf{q}}\cdot\hat{\mathbf{k}}$.

  \subsection{Expansion of the three-halo kernels}
   For our configuration, the arguments of the second-order density kernel given by Equation~\eqref{eq:F2kernel} are
   \begin{equation}
       \mu_{12} = \frac{\mu-\epsilon}{\sqrt{1-2\epsilon\mu+\epsilon^2}}, \quad\quad \frac{k_2}{k_1} = \frac{1}{\epsilon}\sqrt{1-2\epsilon\mu+\epsilon^2}, \quad\quad
       \frac{k_1}{k_2} = \frac{\epsilon}{\sqrt{1-2\epsilon\mu+\epsilon^2}}.
   \end{equation}
   Expanding the denominator and collecting terms gives
   \begin{equation} \label{eq:F2a}
       F_2(\mathbf{q}, \mathbf{k}-\mathbf{q}) = \frac{\mu}{2\epsilon} + \frac{3}{14} + \frac{2\mu^2}{7} + \frac{\epsilon\mu}{14}(8\mu^2-1) + \mathcal{O}(\epsilon^2),
   \end{equation}
   \begin{equation} \label{eq:F2b}
       F_2(-\mathbf{k},\mathbf{q}) = -\frac{\mu}{2\epsilon} + \frac{5}{7} + \frac{2\mu^2}{7} - \frac{\epsilon\mu}{2} + \mathcal{O}(\epsilon^2),
   \end{equation}
   \begin{equation} \label{eq:F2c}
       F_2(\mathbf{k}-\mathbf{q},-\mathbf{k}) = \epsilon^2 \left(\frac{3}{14} - \frac{5\mu^2}{7}\right) + \mathcal{O}(\epsilon^3).
   \end{equation}
   Note the cancellation of the $\mathcal{O}(1/\epsilon)$ poles between Equations~\eqref{eq:F2a} and \eqref{eq:F2b}, as required for a well-defined squeezed limit. The third kernel starts at $\mathcal{O}(\epsilon^2)$ and does not contribute at the order we work, so only two of the three permutations in Equation~\eqref{eq:B_PT_def} survive:
   \begin{equation} \label{eq:BPT_two_terms}
       B^{\rm PT}(\mathbf{q},\mathbf{k}-\mathbf{q},-\mathbf{k}) = 2\,P_{\rm lin}(q)\left[F_2(\mathbf{q}, \mathbf{k}-\mathbf{q})\,P_{\rm lin}(|\mathbf{k}-\mathbf{q}|) + F_2(-\mathbf{k},\mathbf{q})\,P_{\rm lin}(k)\right] + \mathcal{O}(\epsilon^2).
   \end{equation}

   The presence of the $\mu/(2\epsilon)$ pole in Equation~\eqref{eq:F2a} means that $P_{\rm lin}(|\mathbf{k}-\mathbf{q}|)$ must be expanded to second order in $\epsilon$: the $\mathcal{O}(\epsilon^2)$ term multiplies the pole and therefore contributes at $\mathcal{O}(\epsilon)$, which is the order at which the surviving coefficients $A_1$ and $A_3$ appear in Equation~\eqref{eq:B_PT_config}. Writing
   \begin{equation} \label{eq:u_expansion}
       u \equiv \ln\frac{|\mathbf{k}-\mathbf{q}|}{k} = \tfrac{1}{2}\ln\left(1-2\epsilon\mu+\epsilon^2\right) = -\epsilon\mu + \frac{\epsilon^2}{2}\left(1-2\mu^2\right) + \mathcal{O}(\epsilon^3),
   \end{equation}
   and expanding $\ln P_{\rm lin}$ about $\ln k$,
   \begin{equation}
       \ln P_{\rm lin}(|\mathbf{k}-\mathbf{q}|) = \ln P_{\rm lin}(k) + n_k u + \frac{\alpha_k}{2}u^2 + \mathcal{O}(u^3),
       \quad\quad \alpha_k \equiv \frac{\mathrm{d}^2\ln P_{\rm lin}(k)}{\mathrm{d}(\ln k)^2},
   \end{equation}
   exponentiation gives, using $u^2 = \epsilon^2\mu^2 + \mathcal{O}(\epsilon^3)$,
   \begin{equation} \label{eq:Plin_expansion}
       P_{\rm lin}(|\mathbf{k}-\mathbf{q}|) = P_{\rm lin}(k)\left[1 - \epsilon\mu\, n_k + \epsilon^2 c_2\right] + \mathcal{O}(\epsilon^3),
       \quad\quad
       c_2 \equiv \frac{n_k}{2}\left(1-2\mu^2\right) + \frac{n_k^2 + \alpha_k}{2}\,\mu^2.
   \end{equation}
   The $n_k^2$ in $c_2$ arises from the second-order term of the exponential and not from the expansion of $\ln P_{\rm lin}$, which contributes only $\alpha_k$.

   Substituting Equations~\eqref{eq:F2a}, \eqref{eq:F2b}, and \eqref{eq:Plin_expansion} into Equation~\eqref{eq:BPT_two_terms} and collecting powers of $\epsilon$ then gives Equation~\eqref{eq:B_PT_config} with the coefficients of Equation~\eqref{eq:A_i}.

   Inserting Equation~\eqref{eq:B_PT_config} into Equation~\eqref{eq:foreground_pk} and expanding the biased profiles as
   \begin{equation}
       \bU{\rm g}(|\mathbf{k}-\mathbf{q}|) = \bU{\rm g}(k) - q\mu \frac{\partial \bU{\rm g}(k)}{\partial k} + \dots
   \end{equation}
   gives
   \begin{align}
       P_{\Pi_E F}^{\rm 3h}(k_\perp) = \bU{F}(k) \, P_{\rm lin}(k) &\int \frac{\mathrm{d}^3 q}{(2\pi)^3} \, \frac{\mu}{q} \, \Theta(q) \, \bU{\rm g}(q) \, \, P_{\rm lin}(q) \nonumber \\[0.2em]
       &\times \left[A_0 + A_2\mu^2 + \epsilon(A_1 \mu + A_3 \mu^3)\right] \left[\bU{\rm g}(k) - q\mu \partial_k \bU{\rm g}(k)\right].
   \end{align}
   The $\mathcal{O}(\epsilon^0)$ term is even in $\mu$, so the full angular integrand is odd and vanishes. At $\mathcal{O}(\epsilon)$, we obtain Equation~\eqref{eq:P_3h_1}.

   As a sanity check, this calculation recovers the matter-only squeezed-limit response of \cite{2212.11940}, their Equation~(2.9):
   \begin{equation}
       B(\mathbf{q},\mathbf{k}_1,\mathbf{k}_2) \simeq \left[\frac{\mathbf{k}_1 \cdot \mathbf{q}}{q^2} + \frac{10}{7} + \frac{4}{7}\big(\hat{\mathbf{q}}\cdot\hat{\mathbf{k}}_1\big)^2\right] P_{\rm lin}(q) \, P_{\rm lin}(k_1) + (1 \leftrightarrow 2).
   \end{equation}
   Specialising to $\mathbf{k}_1 = \mathbf{k}-\mathbf{q}$, $\mathbf{k}_2 = -\mathbf{k}$ for our configuration and expanding to first order in $\epsilon$,
   \begin{equation}
       B(\mathbf{q},\mathbf{k}-\mathbf{q},-\mathbf{k}) = \left[\frac{13}{7} + \left(\frac{8}{7}-n_k\right)\mu^2 + \mathcal{O}(\epsilon)\right] P_{\rm lin}(q) \, P_{\rm lin}(k),
   \end{equation}
   which recovers the zeroth-order coefficients, $A_0$ and $A_2$, from Equation~\eqref{eq:A_i}. Note that the $\mathcal{O}(\epsilon)$ coefficients, $A_1$ and $A_3$, are not directly tested by this comparison.
   
  \subsection{Expansion of the two-halo kernels}
   The three pieces of the two-halo bispectrum in our configuration are
   \begin{align}
       B^{\rm 2h}_{\mathrm{gg}F}(\mathbf{q}, \mathbf{k}-\mathbf{q}, -\mathbf{k}) &= I_{\rm gg}^1(q, |\mathbf{k}-\mathbf{q}|) \, I_F^1(k) \, P_{\rm lin}(k) \nonumber \\
       &+ I_{\mathrm{g}F}^1(|\mathbf{k}-\mathbf{q}|, k) \, I_{\rm g}^1(q) \, P_{\rm lin}(q) \nonumber \\
       &+ I_{F\mathrm{g}}^1(k,q) \, I_{\mathrm{g}}^1(|\mathbf{k}-\mathbf{q}|) \, P_{\rm lin}(|\mathbf{k}-\mathbf{q}|).
   \end{align}
   For the first, expanding the second argument as
   \begin{equation}
       I_{\rm gg}^1(q, |\mathbf{k}-\mathbf{q}|) = I_{\rm gg}^1(q,k) - q\mu \frac{\partial I_{\rm gg}^1(q,k)}{\partial k} + \dots,
   \end{equation}
   we obtain
   \begin{equation}
       P_{\Pi_E F}^{\mathrm{2h}, (\mathrm{gg}|F), (1)} = - \frac{I_F^1(k) \, P_{\rm lin}(k)}{6\pi^2} \int_0^{k_L} \mathrm{d}q \, q^2 \, \partial_k I_{\rm gg}^1(q,k).
   \end{equation}
   For the second, expanding in the first argument,
   \begin{equation}
       I_{\mathrm{g}F}^1(|\mathbf{k}-\mathbf{q}|, k) = I_{\mathrm{g}F}^1(k,k) - q\mu \left.\frac{\partial I_{\mathrm{g}F}^1(k_1,k_2)}{\partial k_1}\right|_{k_1=k_2=k} + \dots,
   \end{equation}
   \begin{equation}
       P_{\Pi_E F}^{\mathrm{2h}, (\mathrm{g}F|\mathrm{g}), (1)} = - \frac{1}{6\pi^2} \left.\frac{\partial I_{\mathrm{g}F}^1(k_1,k_2)}{\partial k_1}\right|_{k_1=k_2=k} \int_0^{k_L} \mathrm{d}q \, q^2 \, I_{\rm g}^1(q) \, P_{\rm lin}(q).
   \end{equation}
   For the third, both short-scale factors must be expanded,
   \begin{equation}
       I_{\rm g}^1(|\mathbf{k}-\mathbf{q}|) = I_{\rm g}^1(k) - q\mu\partial_k I_{\rm g}^1(k) + \dots, \quad
       P_{\rm lin}(|\mathbf{k}-\mathbf{q}|) = P_{\rm lin}(k)\left[1-\frac{q}{k}\mu n_k\right] + \dots,
   \end{equation}
   giving
   \begin{equation}
       P_{\Pi_E F}^{\mathrm{2h}, (F\mathrm{g}|\mathrm{g}), (1)} = - \frac{P_{\rm lin}(k)}{6\pi^2} \int_0^{k_L} \mathrm{d}q \, q^2 \, I_{F\mathrm{g}}^1(k,q) \left[\frac{n_k}{k} I_{\mathrm{g}}^1(k) + \partial_k I_{\mathrm{g}}^1(k)\right].
   \end{equation}
   The zeroth-order term of each vanishes, and summing the three terms recovers Equation~\eqref{eq:P_2h_1}.

 \section{Spin-1 harmonic decomposition} \label{app:spin-1}
  This appendix gives the harmonic algebra underlying Section~\ref{ssec:full_sky.spin-1}. Throughout we use the spin-raising and lowering operators, $\eth$ and $\bar{\eth}$, the identity $\hat{\mathbf{e}}_{\perp}=\hat{\mathbf{e}}_\theta + i\hat{\mathbf{e}}_\varphi = -\eth\nv$, the plane-wave expansion, and the decomposition
  \begin{equation} \label{eq:nhat_dot_khat}
      \nv\cdot\hat{\mathbf{k}} = \frac{4\pi}{3} \sum_{M=-1}^1 Y_{1M}(\nv)\, Y_{1M}^*(\hat{\mathbf{k}}).
  \end{equation}

  Integrating by parts and substituting the plane-wave expansion, the spin-1 coefficient of the longitudinal piece requires the action of $\eth$ on a product of spin-weighted harmonics,
  \begin{equation}
      \eth\left[Y_{\ell'm'}\, {}_1 Y_{\ell m}^*\right] = (-1)^{m+1} \, \left[ \sqrt{\ell'(\ell'+1)} \, {}_1 Y_{\ell'm'} \, {}_{-1} Y_{\ell, -m} + \sqrt{\ell(\ell+1)} \, Y_{\ell'm'} \, Y_{\ell,-m} \right],
  \end{equation}
  where we have used ${}_1Y_{\ell m}^* = (-1)^{m+1} {}_{-1}Y_{\ell,-m}$. Together with Equation~\eqref{eq:nhat_dot_khat}, this reduces the angular integrals to Gaunt integrals of three spin-weighted harmonics,
  \begin{equation}
     \int\mathrm{d}\nv\, Y_{1M}\, {}_1 Y_{\ell'm'}\, {}_{-1} Y_{\ell,-m} = \sqrt{\frac{3(2\ell'+1)(2\ell+1)}{4\pi}}
     \begin{pmatrix}
         \ell' & \ell & 1 \\
         m' & -m & M
     \end{pmatrix}
     \begin{pmatrix}
         \ell' & \ell & 1 \\
         -1 & 1 & 0
     \end{pmatrix},
  \end{equation}
  \begin{equation}
    \int\mathrm{d}\nv\, Y_{1M}\, Y_{\ell'm'}\, Y_{\ell,-m} = \sqrt{\frac{3(2\ell'+1)(2\ell+1)}{4\pi}}
     \begin{pmatrix}
         \ell' & \ell & 1 \\
         m' & -m & M
     \end{pmatrix}
     \begin{pmatrix}
         \ell' & \ell & 1 \\
         0 & 0 & 0
     \end{pmatrix},
  \end{equation}
  yielding
  \begin{equation}\label{eq:ref_pipar_inter}
      \pi_{\ell m}^{\rm L} = \int\mathrm{d}\chi\,W_{\pi}(\chi) \int\frac{\mathrm{d}^3 k}{(2\pi)^3}\, q^{\rm L}(\mathbf{k},\chi)  \sum_{\ell'm'} i^{\ell'} j_{\ell'}(k\chi) \, \mathcal{W}_{\ell m, \ell' m'}^{\rm L} \, Y_{\ell' m'}^*(\hat{\mathbf{k}}) \, Y_{1,m-m'}^*(\hat{\mathbf{k}}),
  \end{equation}
  \begin{align}
      \mathcal{W}_{\ell m, \ell' m'}^{\rm L} \equiv -\frac{16\pi}{3} (-1)^{m+1} \, &\sqrt{\frac{3(2\ell'+1)(2\ell+1)}{4\pi}}
     \begin{pmatrix}
         \ell' & \ell & 1 \\
         m' & -m & m-m'
     \end{pmatrix} \nonumber \\[0.5em]
     \times &\left[\sqrt{\ell'(\ell'+1)}
     \begin{pmatrix}
        \ell' & \ell & 1 \\
         -1 & 1 & 0
     \end{pmatrix}
     + \sqrt{\ell(\ell+1)}
     \begin{pmatrix}
         \ell' & \ell & 1 \\
         0 & 0 & 0
     \end{pmatrix}\right].
  \end{align}
  The selection rules of the Wigner-$3j$ symbols enforce that $\ell'=\ell\pm1$. Moreover,
   in the frame $\hat{\mathbf{k}}=\hat{\mathbf{z}}$, the product $Y_{\ell' m'}^*(\hat{\mathbf{k}}) \, Y_{1,m-m'}^*(\hat{\mathbf{k}})$ forces $m'=m=0$.

  Computing $\bar{\pi}_{\ell m}^{\rm L}$ in an analogous way, with $\bar{\eth}$ in place of $\eth$ and $(q^{\rm L}(\mathbf{k}))^* = q^{\rm L}(-\mathbf{k})$, produces the same Wigner structure with the lower rows of the second $3j$ symbol permuted. Using the sign-reversal identity, the extra phase is $(-1)^{\ell'+\ell+1}$, which is 1 whenever the $3j$ symbol with all $m=0$ in the bottom row is non-zero. The remaining phase difference is cancelled by the transformation of the complex-conjugated field under $\mathbf{k} \to -\mathbf{k}$. Hence, $\bar{\pi}_{\ell m}^{\rm L} = \pi_{\ell m}^{\rm L}$ and, from Equation~\eqref{eq:EB_full_sky_def}, it follows that $\pi_{\ell m}^{{\rm L},B}=0$.

  The final result requires an explicit calculation of the remaining Wigner-$3j$ symbols. This can be achieved by exploiting their relation with the Clebsch--Gordan coefficients. The relevant symbols are:
  \begin{equation}
    \wtj{\ell}{1}{\ell'}{m}{-m-m'}{m'}=\frac{(-1)^{\ell'}}{\sqrt{2\ell+1}}f^{\ell'}_{m,m'},
  \end{equation}
  where
  \begin{align}
    &f^{\ell+1}_{0,0}=\sqrt{\frac{\ell+1}{2\ell+3}},\hspace{12pt}
    f^{\ell-1}_{0,0}=-\sqrt{\frac{\ell}{2\ell-1}},\\
    &f^\ell_{\pm1,0}=\mp\frac{1}{\sqrt{2}},\hspace{12pt}
    f^{\ell+1}_{\pm1,0}=\sqrt{\frac{\ell}{2(2\ell+3)}},\hspace{12pt}
    f^{\ell-1}_{\pm1,0}=\sqrt{\frac{\ell}{2(2\ell-1)}},\\
    &f^\ell_{1,-1}=\frac{1}{\sqrt{\ell(\ell+1)}},\hspace{12pt}
    f^{\ell+1}_{1,-1}=-\sqrt{\frac{\ell(\ell+2)}{(2\ell+3)(\ell+1)}},\hspace{12pt}
    f^{\ell-1}_{1,-1}=\sqrt{\frac{(\ell+1)(\ell-1)}{\ell(2\ell-1)}}.
  \end{align}
  Replacing these in Equation~\eqref{eq:ref_pipar_inter}, and using the following recursion relations for the spherical Bessel functions
  \begin{align}\label{eq:bessel_rec_1}
    &j_{\ell-1}(x) + j_{\ell+1}(x) = \frac{2\ell+1}{x} \, j_{\ell}(x),\\\label{eq:bessel_rec_2}
    &\ell j_{\ell-1}(x)-(\ell+1)j_{\ell+1}(x)=(2\ell+1)\,j'_\ell(x)
  \end{align}
  we obtain the final result given by Equation~\eqref{eq:pi_lm_par_E}.

  We now consider the transverse component. Choosing $\hat{\mathbf{k}}=\hat{\mathbf{z}}$ and $\mathbf{q}^{\rm T} = q_\perp \hat{\mathbf{y}}$, we have
  \begin{equation}
      \nv\cdot\mathbf{q}^{\rm T} = iq_\perp \sqrt{\frac{2\pi}{3}}\left[Y_{1,1}(\nv) + Y_{1,-1}(\nv)\right],
  \end{equation}
  so only the modes $M=\pm1$ contribute. Since $Y_{\ell'm'}^*(\hat{\mathbf{z}})$ enforces $m'=0$, the selection rule $M=m-m'$ gives $m=\pm1$. This is in contrast to the longitudinal component, which is supported only at $m=0$.

  Then, following steps analogous to the longitudinal case, we obtain
  \begin{equation}\nonumber
      \pi_{\ell m}^{\rm T} = \delta_{m,\pm1} i^{\ell+1} \sqrt{4\pi(2\ell+1)} \int\frac{\mathrm{d}\chi}{2}\, W_\pi(\chi) \int\frac{\mathrm{d}^3 k}{(2\pi)^3}\, \mathbf{q}_\perp(\mathbf{k},\chi)\, \left[j_\ell'(k\chi) + \frac{j_\ell(k\chi)}{k\chi} \mp i j_\ell(k\chi)\right],
  \end{equation}
  \begin{equation}\nonumber
      \bar{\pi}_{\ell m}^{\rm T} = \delta_{m,\pm1} i^{\ell+1} \sqrt{4\pi(2\ell+1)} \int\frac{\mathrm{d}\chi}{2}\, W_\pi(\chi) \int\frac{\mathrm{d}^3 k}{(2\pi)^3}\, \mathbf{q}_\perp(\mathbf{k},\chi)\, \left[j_\ell'(k\chi) + \frac{j_\ell(k\chi)}{k\chi} \pm i j_\ell(k\chi)\right].
  \end{equation}
  Combining these results according to Equation~\eqref{eq:EB_full_sky_def}, we obtain Equations~\eqref{eq:pi_lm_perp_E} and \eqref{eq:pi_lm_perp_B} for the $E$- and $B$-modes of the transverse component.

 \section{The radial velocity contribution} \label{app:radial}
  The moving lens effect is sourced only by the components of the velocity and potential gradient transverse to the line of sight, whereas Equation~\eqref{eq:lambda} and everything that follows from it are written in terms of the full scalar product. The two differ by a radial term,
  \begin{equation}
      \mathbf{v}_\perp \cdot \boldsymbol{\nabla}_\perp \phi = \mathbf{v} \cdot \boldsymbol{\nabla}\phi - v_\parallel(\nabla_\parallel\phi).
  \end{equation}
  This appendix derives the radial contribution in full and establishes the result quoted in Section~\ref{ssec:ml.definition} that its contribution to the cross-correlation with the projected galaxy momentum vanishes at leading order in the squeezed limit and is suppressed by $\mathcal{O}(q^2/k^2)$ beyond it. 

  \subsection{Harmonic decomposition of the radial source}
   We write the radial contribution in the same form as Equation~\eqref{eq:Lambda}, so that the two may be compared directly:
   \begin{equation} \label{eq:lambda_r_def}
       \lambda_r(\nv) = \int\mathrm{d}\chi\, W_\lambda(\chi)\, \Lambda_r(\chi\nv), \quad\quad \Lambda_r(\mathbf{k}) = \int_{\mathbf{k}}\mathrm{D}k_1\,\mathrm{D}k_2\, \frac{(\mathbf{k}_1\cdot\nv)(\mathbf{k}_2\cdot\nv)}{k_1^2 k_2^2}\, \delta_{v}(\mathbf{k}_1)\,\delta_{\rm m}(\mathbf{k}_2).
   \end{equation}
   This is simply Equation~\eqref{eq:Lambda} with $\mathbf{k}_1 \cdot \mathbf{k}_2$ replaced by the product of the two radial projections. The ML source is then given by $\Lambda - \Lambda_r$.

   In configuration space, $\Lambda_r$ is generated by $\partial_\chi\phi_v\,\partial_\chi\phi$, where $\phi_v$ is the velocity potential, sourced by $\delta_v$, and $\phi$ is the gravitational potential of Equation~\eqref{eq:lambda}, sourced by $\delta_{\rm m}$. Integrating by parts and neglecting boundary terms,
   \begin{equation}
       \lambda_r(\nv) = \int\mathrm{d}\chi\,\partial_\chi\phi_v\,\partial_\chi\phi = -\int\mathrm{d}\chi\,\phi_v\,\partial_\chi^2\phi,
   \end{equation}
   and, using $\partial_\chi = \hat{n}_i \partial_i$,
   \begin{equation}
       \lambda_r(\nv) = -\int\mathrm{d}\chi\,\phi_v(\mathbf{x})\, \hat{n}_i \hat{n}_j\, \partial_i \partial_j \phi(\mathbf{x}), 
   \end{equation}
   where $\mathbf{x} \equiv \chi\nv$. Expanding the fields in Fourier space and applying the convolution theorem,
   \begin{equation}
       (\phi_v\,\partial_i\partial_j\phi)(\mathbf{k}) = -\int\frac{\mathrm{d}^3q}{(2\pi)^3}\, \,q_i\, q_j\, \phi_v(\mathbf{k}-\mathbf{q})\, \phi(\mathbf{q}),
   \end{equation}
   so that, restoring the normalisation of Equation~\eqref{eq:lambda_r_def},
   \begin{equation}
       \lambda_r(\nv) = -\int\mathrm{d}\chi\,W_\lambda(\chi) \int\frac{\mathrm{d}^3 k}{(2\pi)^3} \int\frac{\mathrm{d}^3q}{(2\pi)^3}\, (\hat{\mathbf{q}}\cdot\nv)^2\, \frac{\delta_{\rm m}(\mathbf{q})\,\delta_{v}(\mathbf{k}-\mathbf{q})}{|\mathbf{k}-\mathbf{q}|^2}\, e^{i\mathbf{k}\cdot(\chi\nv)}.
   \end{equation}
   Two features of this expression drive everything that follows. First, all angular dependence is carried by $(\hat{\mathbf{q}}\cdot\nv)^2$, which is quadratic and therefore even. Second, one power of $q$ has been traded for a power of $|\mathbf{k}-\mathbf{q}|$ relative to Equation~\eqref{eq:Lambda}, so the radial term is already smaller by $\mathcal{O}(q/k)$ in the squeezed limit before any angular averaging is performed.

   Using the plane wave expansion, the harmonic coefficients are
   \begin{equation}\nonumber
       \lambda_{r,\ell m} = -4\pi \sum_{\ell'm'} i^{\ell'} \int\mathrm{d}\chi\, W_\lambda(\chi) \int\frac{\mathrm{d}^3 k}{(2\pi)^3}\, j_{\ell'}(k\chi) \, Y_{\ell'm'}^*(\hat{\mathbf{k}}) \int\frac{\mathrm{d}^3 q}{(2\pi)^3}\, \frac{\delta_{\rm m}(\mathbf{q})\,\delta_{v}(\mathbf{k}-\mathbf{q})}{|\mathbf{k}-\mathbf{q}|^2} \, A_{\ell m,\ell'm'}(\hat{\mathbf{q}}),
   \end{equation}
   where
   \begin{equation}
       A_{\ell m,\ell'm'}(\hat{\mathbf{q}}) \equiv \int\mathrm{d}\nv\, Y_{\ell m}^*(\nv)\, Y_{\ell'm'}(\nv)\, (\nv\cdot\hat{\mathbf{q}})^2.
   \end{equation}
   Writing the angular dependence in terms of the Legendre polynomial of degree two,
   \begin{equation}
       (\nv\cdot\hat{\mathbf{q}})^2 = \frac{1}{3} + \frac{2}{3} P_2(\nv\cdot\hat{\mathbf{q}}),
   \end{equation}
   splits $A$ into two contributions $A = A^{(0)} + A^{(2)}$, separating the harmonic coefficients into monopole and quadrupole pieces, $\lambda_{r,\ell m} = \lambda_{r,\ell m}^{(0)} + \lambda_{r,\ell m}^{(2)}$. As in the main text, we adopt the frame $\hat{\mathbf{k}}=\hat{\mathbf{z}}$.

   For the monopole, orthogonality of the spherical harmonics gives 
   \begin{equation}
       A_{\ell m,\ell'm'}^{(0)} = \frac{1}{3} \delta_{\ell\ell'} \delta_{mm'},
   \end{equation}
   and, since $Y_{\ell m}^*(\hat{\mathbf{z}}) = \sqrt{(2\ell+1)/(4\pi)}\,\delta_{m,0}$ in this frame, only axisymmetric components contribute:
   \begin{equation}
       \lambda_{r,\ell m}^{(0)} = -\frac{1}{3} i^{\ell} \delta_{m,0} \sqrt{4\pi(2\ell+1)} \int\mathrm{d}\chi\, W_\lambda(\chi)\int\frac{\mathrm{d}^3 k}{(2\pi)^3}\, j_{\ell}(k\chi) \int\frac{\mathrm{d}^3 q}{(2\pi)^3}\, \frac{\delta_{\rm m}(\mathbf{q})\,\delta_{v}(\mathbf{k}-\mathbf{q})}{|\mathbf{k}-\mathbf{q}|^2}.
   \end{equation}

   For the quadrupole,
   \begin{equation}
       A_{\ell m,\ell'm'}^{(2)} = \frac{2}{3}\int\mathrm{d}\nv\, Y_{\ell m}^*(\nv)\, Y_{\ell'm'}(\nv)\, P_2(\nv\cdot\hat{\mathbf{q}}) = \frac{8\pi}{15}\sum_M Y_{2M}^*(\hat{\mathbf{q}}) \int\mathrm{d}\nv\, Y_{\ell m}^*\, Y_{\ell'm'}\, Y_{2M},
   \end{equation}
   where we have used the addition theorem $P_2(\nv\cdot\hat{\mathbf{q}}) = \frac{4\pi}{5}\sum_M Y_{2M}(\nv)\, Y_{2M}^*(\hat{\mathbf{q}})$. The remaining integral is a Gaunt integral,
   \begin{equation}
       \int\mathrm{d}\nv\, Y_{\ell m}^* Y_{\ell'm'} Y_{2M} = (-1)^m \sqrt{\frac{5(2\ell+1)(2\ell'+1)}{4\pi}}
       \begin{pmatrix}
           \ell & \ell' & 2 \\
           0 & 0 & 0
       \end{pmatrix}
       \begin{pmatrix}
           \ell & \ell' & 2 \\
           -m & m' & M
       \end{pmatrix}.
   \end{equation}
   The Wigner-$3j$ symbols impose $|\ell-\ell'| \leq 2$ with $\ell+\ell'+2$ even, so only $\ell'\in\{\ell-2,\ell,\ell+2\}$ contribute. The quadrupole therefore introduces couplings between multipoles separated by $\Delta\ell=2$, in contrast to the monopole, which is diagonal. The frame choice $\hat{\mathbf{k}}=\hat{\mathbf{z}}$ forces $m'=0$, and the second Wigner symbol then fixes $M=m$, isolating a single harmonic $Y_{2m}^*(\hat{\mathbf{q}})$. Only $m=0$ can correlate with a scalar field, and for that component the two Wigner symbols coincide. Using $Y_{20}(\hat{\mathbf{q}}) = \sqrt{5/(16\pi)}\,(3\mu^2-1)$ with $\mu \equiv \hat{\mathbf{q}}\cdot\hat{\mathbf{k}}$, the quadrupole carries the same overall prefactor as the monopole piece, multiplying
   \begin{equation} \label{eq:quad_sum}
       (3\mu^2-1) \sum_{\ell'} i^{\ell'} (2\ell'+1)
       \begin{pmatrix}
           \ell & \ell' & 2 \\
           0 & 0 & 0
       \end{pmatrix}^2
       j_{\ell'}(k\chi).
   \end{equation}
   Note that the harmonic formalism has converted the line-of-sight angle $\hat{\mathbf{q}}\cdot\nv$ of the original expression into the angle $\mu$ between the soft mode and $\hat{\mathbf{k}}$. This is what makes the comparison with the transverse signal possible, since the latter is controlled by the same $\mu$.

   The three terms in Equation~\eqref{eq:quad_sum} are not independent. Defining 
   \begin{equation}
       \mathcal{D}_{\ell,\ell'} \equiv \mathcal{N}_\ell\, (2\ell'+1)\,
       \begin{pmatrix}
           \ell & \ell' & 2 \\
           0 & 0 & 0
       \end{pmatrix}^2,
       \quad\quad
       \mathcal{N}_{\ell} \equiv \sqrt{\frac{8\pi(2\ell+1)}{3}},
   \end{equation}
   the relevant Wigner-$3j$ symbols give
   \begin{equation} \label{eq:Dll'_over_Nl}
       \frac{\mathcal{D}_{\ell,\ell-2}}{\mathcal{N}_\ell} = \frac{3\ell(\ell-1)}{2(2\ell-1)(2\ell+1)}, \quad \frac{\mathcal{D}_{\ell,\ell}}{\mathcal{N}_\ell} = \frac{\ell(\ell+1)}{(2\ell-1)(2\ell+3)}, \quad \frac{\mathcal{D}_{\ell,\ell+2}}{\mathcal{N}_\ell} = \frac{3(\ell+1)(\ell+2)}{2(2\ell+1)(2\ell+3)}.
   \end{equation}
   Applying the recurrence relation Equation~\eqref{eq:bessel_rec_2} twice yields
   \begin{align}
       j_\ell''(x) &= \frac{\ell(\ell-1)}{(2\ell-1)(2\ell+1)}\,j_{\ell-2}(x) + \frac{(\ell+1)(\ell+2)}{(2\ell+1)(2\ell+3)}\,j_{\ell+2}(x) \nonumber \\[0.5em]
       &- \left[\frac{\ell^2}{(2\ell-1)(2\ell+1)} + \frac{(\ell+1)^2}{(2\ell+1)(2\ell+3)}\right]\,j_\ell(x). \label{eq:jl_pp}
   \end{align}
   Comparing with Equation~\eqref{eq:Dll'_over_Nl}, the $\ell'=\ell\pm2$ coefficients are exactly $3/2$ times the corresponding coefficients in Equation~\eqref{eq:jl_pp}. Restoring the $i^{\ell'}$ prefactor from the plane wave expansion, the sum over $\ell'$ collapses to
   \begin{equation}
       \sum_{\ell'=\ell,\ell\pm2} i^{\ell'}\, \mathcal{D}_{\ell,\ell'}\, j_{\ell'}(x) = -\frac{i^\ell}{2}\sqrt{\frac{8\pi(2\ell+1)}{3}}\left[3j_{\ell}''(x) + j_{\ell}(x)\right].
   \end{equation}
   Adding the monopole and quadrupole, the $j_\ell$ terms combine, and the radial contribution reduces to the single expression
   \begin{align}\label{eq:lambda_r_lm}
       \lambda_{r,\ell m} = -\frac{1}{2}\,\delta_{m,0}\, i^\ell \sqrt{4\pi(2\ell+1)} \int\mathrm{d}\chi\, W_\lambda(\chi) \int\frac{\mathrm{d}^3k}{(2\pi)^3} &\int\frac{\mathrm{d}^3 q}{(2\pi)^3}\, \frac{\delta_{\rm m}(\mathbf{q})\,\delta_{v}(\mathbf{k}-\mathbf{q})}{|\mathbf{k}-\mathbf{q}|^2}  \\[0.2em]\nonumber
       \times &\left[(1-\mu^2)\,j_{\ell}(k\chi) - (3\mu^2-1)\,j_{\ell}''(k\chi)\right]. 
   \end{align}
   Equation~\eqref{eq:lambda_r_lm} has the same structure as the scalar coefficients of Equation~\eqref{eq:lambda_lm}. The radial contribution therefore behaves exactly like an additional scalar field on the sky, and can be characterised by an effective source in place of $\Lambda$ since the $j_{\ell}''$ contribution is strongly suppressed in the Limber approximation at high $\ell$.

   Dropping the $j_{\ell}''$ term, it is convenient to define
   \begin{equation}
       \Lambda_2(\mathbf{k}) \equiv -\Lambda_r(\mathbf{k}) = \frac{1}{2}\int\frac{\mathrm{d}^3 q}{(2\pi)^3}\, \frac{1-\mu^2}{|\mathbf{k}-\mathbf{q}|^2}\, \delta_{v}(\mathbf{q})\, \delta_{\rm m}(\mathbf{k}-\mathbf{q}),
   \end{equation}
   where we have relabelled $\mathbf{q}\to\mathbf{k}-\mathbf{q}$ so that the velocity field has the same leg as in Equation~\eqref{eq:Lambda}. This is permitted because the kernel is invariant under this change of variables.
   
   Writing $\Lambda_1 \equiv \Lambda$ for the full scalar product of Equation~\eqref{eq:Lambda}, the ML signal, which is sourced only by the transverse components, follows from $\Lambda_1 + \Lambda_2$. The two kernels may then be compared directly:
   \begin{equation} \label{eq:K1_K2}
      K_1(\mathbf{q},\mathbf{k}-\mathbf{q}) = \frac{\mathbf{q}\cdot(\mathbf{k}-\mathbf{q})}{q^2|\mathbf{k}-\mathbf{q}|^2}, \quad\quad
      K_2(\mathbf{q},\mathbf{k}-\mathbf{q}) = \frac{1}{2}\,\frac{q^2 - (\mathbf{q}\cdot\hat{\mathbf{k}})^2}{q^2\,|\mathbf{k}-\mathbf{q}|^2}.
   \end{equation}

  \subsection{Cross-spectrum with the galaxy momentum}
   Both $K_1$ and $K_2$ are symmetric under $\mathbf{q}\leftrightarrow\mathbf{k}-\mathbf{q}$. For $K_1$ this is immediate; for $K_2$ it follows from a useful identity. Writing $\mu'$ for the angle between $\mathbf{k}-\mathbf{q}$ and $\hat{\mathbf{k}}$,
   \begin{equation}
      \mu' = \frac{\mathbf{k}\cdot(\mathbf{k}-\mathbf{q})}{k|\mathbf{k}-\mathbf{q}|} = \frac{k}{|\mathbf{k}-\mathbf{q}|} - \frac{q}{|\mathbf{k}-\mathbf{q}|}\,\mu,
   \end{equation}
   so that
   \begin{equation} \label{eq:mu_identity}
      \frac{1-\mu'^2}{q^2} = \frac{1-\mu^2}{|\mathbf{k}-\mathbf{q}|^2},
   \end{equation}
   which is the statement that $\mathbf{q}$ and $\mathbf{k}-\mathbf{q}$ have equal and opposite components perpendicular to $\hat{\mathbf{k}}$. The two Wick contractions of $\Lambda_2$ against the galaxy momentum are generated by kernels written in terms of $\mu$ and $\mu'$ respectively, and Equation~\eqref{eq:mu_identity} shows that these are the same function. The kernel therefore factors out of the sum over contractions, exactly as it does for $\Lambda_1$.

   Writing the two contractions explicitly in terms of the spectral combination $\mathcal{S}$ of Equation~\eqref{eq:calS}, the two cross-spectra with the longitudinal momentum of Section~\ref{sec:full_sky} are
   \begin{equation} \label{eq:P_qLambda1_full}
       P_{q^{\rm L} \Lambda_1}(k) = aHf\int\frac{\mathrm{d}^3 q}{(2\pi)^3}\, \Theta(q)\, \frac{\mathbf{q}\cdot\hat{\mathbf{k}}}{q^2}\; \frac{\mathbf{q}\cdot(\mathbf{k}-\mathbf{q})}{q^2|\mathbf{k}-\mathbf{q}|^2}\; \mathcal{S}(\mathbf{q},\mathbf{k}-\mathbf{q}),
   \end{equation}
   \begin{equation} \label{eq:P_qLambda2_full}
      P_{q^{\rm L} \Lambda_2}(k) = aHf\int\frac{\mathrm{d}^3 q}{(2\pi)^3}\, \Theta(q)\, \frac{\mathbf{q}\cdot\hat{\mathbf{k}}}{q^2}\; \frac{1}{2}\,\frac{(1-\mu^2)}{|\mathbf{k}-\mathbf{q}|^2}\; \mathcal{S}(\mathbf{q},\mathbf{k}-\mathbf{q}).
   \end{equation}
   The two expressions differ only in the source kernel, given by Equation~\eqref{eq:K1_K2}. Since $\mathcal{S}$ is common to both, it cancels in the ratio, and the conclusions below hold independent of the relationship between $P_{\mathrm{g}v}$ and $P_{\mathrm{gm}}$. In the squeezed limit, they behave differently. For $\Lambda_1$, $|\mathbf{k}-\mathbf{q}| \to k$ and $\mathbf{q}\cdot(\mathbf{k}-\mathbf{q}) \to qk\mu$, so the two factors of $\mathbf{q}\cdot\hat{\mathbf{k}}$ combine into $\mu^2$, with angular average $\langle \mu^2 \rangle = 1/3$. The result is Equation~\eqref{eq:P_qLambda_sq}, the expression quoted in Section~\ref{sec:full_sky},
   \begin{equation} \label{eq:P_qT1_sq}
      P_{q^{\rm L} \Lambda_1}(k) \simeq \frac{aHf}{3k}\left[P_{\mathrm{gm}}(k) \, \mathcal{A}_v(k_L) + P_{\mathrm{g}v}(k) \, \mathcal{A}_{\rm m}(k_L)\right].
   \end{equation}
   For $\Lambda_2$, the same limit gives $K_2 \to (1-\mu^2)/2k^2$, which is even in $\mu$. Multiplied by the odd factor $\mathbf{q}\cdot\hat{\mathbf{k}}$ from the momentum kernel, the angular integrand is odd overall, and thus the radial contribution vanishes identically in the squeezed limit. This is the same mechanism that removes the foreground bispectrum in Section~\ref{ssec:foregrounds.squeezed_limit}.

   The order of the residual now follows immediately. As noted above, the radial kernel is already suppressed by one power of $q/k$ relative to $K_1$. Had its angular average been non-zero, the radial contribution would therefore have appeared at $\mathcal{O}(q/k)$. Since the leading term instead cancels, the first non-vanishing contribution arises only at the next order in the expansion, yielding
   \begin{equation}
       \frac{P_{q^{\rm L} \Lambda_2}(k)}{P_{q^{\rm L} \Lambda_1}(k)} = \mathcal{O}\left(\frac{q^2}{k^2}\right) \leq \mathcal{O}\left(\frac{k_L^2}{k^2}\right),
   \end{equation}
   where the inequality follows from the reconstruction filter, which enforces $q<k_L$.

   This result has two important consequences. First, the radial residual is suppressed by one additional power of $q/k$ compared with the residual foreground contamination discussed in Section~\ref{ssec:foregrounds.beyond_squeezed}, which enters at $\mathcal{O}(q/k)$. Consequently, throughout the regime in which the foreground analysis is valid, the radial contribution is parametrically negligible. Second, the same reconstruction filter that suppresses foreground contamination also suppresses the radial contribution, but by an additional power of $q/k$. The approximation introduced in Section~\ref{ssec:ml.definition} therefore becomes increasingly accurate as the reconstruction scale $k_L$ is lowered, making it most reliable precisely in the regime for which the estimator is intended.

\bibliographystyle{JHEP}
\bibliography{references}

\end{document}